\documentclass[aps,prx,twocolumn,showpacs,superscriptaddress,longbibliography]{revtex4-2}
\usepackage{amsmath,amssymb,graphicx}
\usepackage{graphicx,epstopdf}
\usepackage{gensymb}
\usepackage[dvipsnames]{xcolor}
\usepackage[T1]{fontenc}
\usepackage[utf8]{inputenc}
\newcommand{\be}{\begin{equation}}
\newcommand{\ee}{\end{equation}}
\newcommand{\bea}{\begin{eqnarray}}
\newcommand{\eea}{\end{eqnarray}}
\newcommand{\bse}{\begin{subequations}}
\newcommand{\ese}{\end{subequations}}

\usepackage{multibib}
\usepackage{color}
\usepackage[colorlinks,bookmarks=false,citecolor=darkblue,linkcolor=red,urlcolor=blue]{hyperref}

\definecolor{darkred}{rgb}{0.7,0.0,0.0}

\definecolor{darkblue}{rgb}{0,0.02,0.45}

\definecolor{darkgreen}{rgb}{0.02,0.45,0.0}

\definecolor{violet}{rgb}{0.8,0.2,0.6}

\begin{document}

\title{Structural and double magnetic transitions in the frustrated spin-$\frac{1}{2}$ capped-kagome antiferromagnet (RbCl)Cu$_{5}$P$_{2}$O$_{10}$}

\author{S. Mohanty}
\affiliation{School of Physics, Indian Institute of Science Education and Research Thiruvananthapuram-695551, India}
\author{J. Babu}
\author{Y. Furukawa}
\affiliation{Ames National Laboratory and Department of Physics and Astronomy, Iowa State University, Ames, Iowa 50011, USA}
\author{R. Nath}
\email{rnath@iisertvm.ac.in}
\affiliation{School of Physics, Indian Institute of Science Education and Research Thiruvananthapuram-695551, India}

\date{\today}

\begin{abstract}
The structural and magnetic properties of the geometrically frustrated spin-$1/2$ capped-kagome antiferromagnet (RbCl)Cu$_{5}$P$_{2}$O$_{10}$ are investigated via temperature dependent x-ray diffraction, magnetization, heat capacity, and $^{31}$P NMR experiments on a polycrystalline sample. It undergoes a structural transition at around $T_{\rm t} \simeq 310$~K from a high temperature trigonal ($P\bar{3}m1$) to a low temperature monoclinic ($C2/c$) unit cell, where the low temperature structure features the capped-kagome geometry of Cu$^{2+}$ ions. Interestingly, it shows the onset of two successive magnetic transitions at $T_{\rm N1} \simeq 20$~K and $T_{\rm N2} \simeq 7$~K. The shape of the $^{31}$P NMR spectra unfold the possible nature of the transitions below $T_{\rm N1}$ and $T_{\rm N2}$ to be of incommensurate and commensurate antiferromagnetic type, respectively. A large value of the Curie-Weiss temperature as compared to $T_{\rm N1}$ sets the frustration parameter $f \simeq 8$, ensuring strong magnetic frustration in the compound. From the $^{31}$P NMR spin-lattice relaxation rate, the leading antiferromagnetic exchange coupling is estimated to be $J/k_{\rm B} \simeq 117$~K. These unusual double magnetic transitions make this compound beguiling for further investigations.
\end{abstract}

%\pacs{ }
\maketitle

\section{Introduction}
Recently, geometrically frustrated magnets have received enormous attention because of their potential to host a rich variety of magnetic and structural phases under different conditions~\cite{Ramirez453,Diep2013}. One of the most engrossing ground states is the quantum spin liquid (QSL), a highly entangled and dynamically disordered state, lacking magnetic long-range-order (LRO)~\cite{Balents199}. Among different frustrated magnets, triangular lattice is the simplest example of a geometrically frustrated magnet where the spins are arranged at the vertices of a triangle, defying the antiparallel arrangements, and leading to magnetic frustration. In a triangular lattice, the triangles are edged shared to form a two-dimensional (2D) layer. While in a kagome lattice, the triangles are corner shared in a 2D layer which amplifies the degree of frustration compared to an edge shared triangular lattice. In the perfect low spin ($S =1/2$) kagome lattice antiferromagnets (KLAFs), quantum fluctuations along with strong magnetic frustration impede the conventional magnetic LRO and gives rise to a strongly correlated QSL or other non-trivial ground states~\cite{Yan1173}. A celebrated example is herbertsmithite ZnCu$_3$(OH)$_6$Cl$_2$, which is the first realization of a perfect $S=1/2$ kagome compound with no magnetic LRO down to 50~mK, confirming QSL~\cite{Han406,Khuntia469,Fu655}.
%though the exact nature of QSL is still under debate~\cite{Helton107204,Han406}.

Several phase diagrams have been reported theoretically for KLAFs, predicting a large variety of ground states as a function of the ratio of exchange couplings and also applied magnetic field~\cite{Bieri060407,Picot060407,Suttner020408}.
%In the presence of applied field, the spin nematic order, super-solid, super-fluid, Valance Bond Crystal (VBC), and QSL phases are expected for the kagome lattice, regardless of the spin value~\cite{Picot060407}.
%\textbf{Moreover, theoretical study of the field induced phase diagram of KHAF proposed several kinds of exotic phases including magnetization plateau~\cite{Okamoto180407}. This phase is corresponding to the collinear spin-liquid phase with degenerate up-up-down (uud) configuration on the triangles.}
Experimentally, many compounds have been discovered and investigated which either show a subtle deviation from ideal 2D kagome geometry or are having strong anisotropy, leading to magnetic LRO and/or other exotic phases at finite temperatures~\cite{Hering10}. Apart from perfect kagome lattice, many special derivatives of kagome lattice have also been discovered that include square-kagome~\cite{Fujihala3429}, octa-kagome~\cite{Tang14057}, staircase-kagome~\cite{Morosan144403,Yoo2397}, sphere-kagome~\cite{Rousochatzakis094420}, strip-kagome~\cite{Jeschke140410,Morita2399}, tripod-kagome~\cite{Dun157201,Dun031069}, hyper-kagome~\cite{Okamoto137207,Lawler227201} etc. Because of the more intricate lattice geometry, these derivatives harbor an array of intriguing magnetic ground states. To name a few, QSL is realized in the square-kagome compound KCu$_6$AlBiO$_4$(SO$_4$)$_5$Cl~\cite{Fujihala3429} and hyperkagome compound Na$_4$Ir$_3$O$_8$~\cite{Okamoto137207}, spin-singlet and antiferromagnetic ordering are observed in octa-kagome lattices BiOCu$_2$(Te,Se)O$_3$(SO$_4$)(OH)·H$_2$O, respectively~\cite{Tang14057}, a spin-ice type ground state is reported in the tripod kagome lattice Mg$_2$Dy$_3$Sb$_3$O$_{14}$~\cite{Dun157201} etc.

Recently, a new kagome variant, capped-kagome lattice compound with general formula ($MX$)Cu$_5$O$_2$($T^{5+}$O$_4$)$_2$ [$M =$ K, Rb, Cs, Cu; $X =$ Cl, Br; $T^{5+}=$ P, V] is being pursued rigorously. This series belongs to the Averievite family, an oxide mineral found as a product of post-eruption volcanic activity. Most of these compounds show a structural transition from a high symmetric trigonal at high temperatures to a low symmetric monoclinic phase at low temperatures. In this series, the magnetic properties of the only compound (CsCl)Cu$_5$V$_2$O$_{10}$ (CCCVO) are elaborately studied which undergoes the structural transition at 310~K~\cite{Botana054421,Dey125106}. Magnetic susceptibility reveals a large Curie-Weiss (CW) temperature $\theta_{\rm CW} \simeq -185$~K and the compound encounters a magnetic LRO at around $T_{\rm N} \simeq 24$~K. Zn$^{2+}$ substitution at the capped Cu$^{2+}$ site [i.e., (CsCl)Cu$_{5-x}$Zn$_x$V$_2$O$_{10}$] suppresses the magnetic as well as the structural transitions and is expected to yield QSL for $x=2$~\cite{Georgopoulou14742,BiesnerL060410}.
%The combined structural and magnetic phase diagram of the Zn$^{2+}$ substituted (CsCl)Cu$_{5-x}$Zn$_x$V$_2$O$_{10}$ ($x = 0$, 0.25, 0.5, 0.75, 1, and 1.25) is explored via time-domain magneto-THz and ESR spectroscopies~\cite{BiesnerL060410}.
%Recently, Zn$^{2+}$ substituted sample with $x = 2$ reveals magnetic excitations compatible with that expected for a QSL and indicative of proximity to the quantum critical point~\cite{Georgopoulou14742}.
%In (CsCl)Cu5V2O10 Ionic radii of P$^{5+}$ is two times smaller than that of V$^{5+}$.
Some preliminary magnetic data are also reported for the phosphate analog compounds (CsCl, CsBr, CsI)Cu$_5$P$_2$O$_{10}$~\cite{Winiarski4328}.
%and the value of $\theta_{\rm CW}$ is found to be increasing systematically as the halide ion is changed from Cl to I in the order of increasing ionic radius~\cite{Winiarski4328}.
Recently, two more new capped-kagome compounds (molybdate-tellurites) based on Ni$^{2+}$ and Cu$^{2+}$ are also reported~\cite{Zhang2299}. Thus, the interesting structural aspects and highly frustrated geometry make these compounds very special in the context of new quantum phase transitions.

\begin{figure*}
\includegraphics[width=\linewidth]{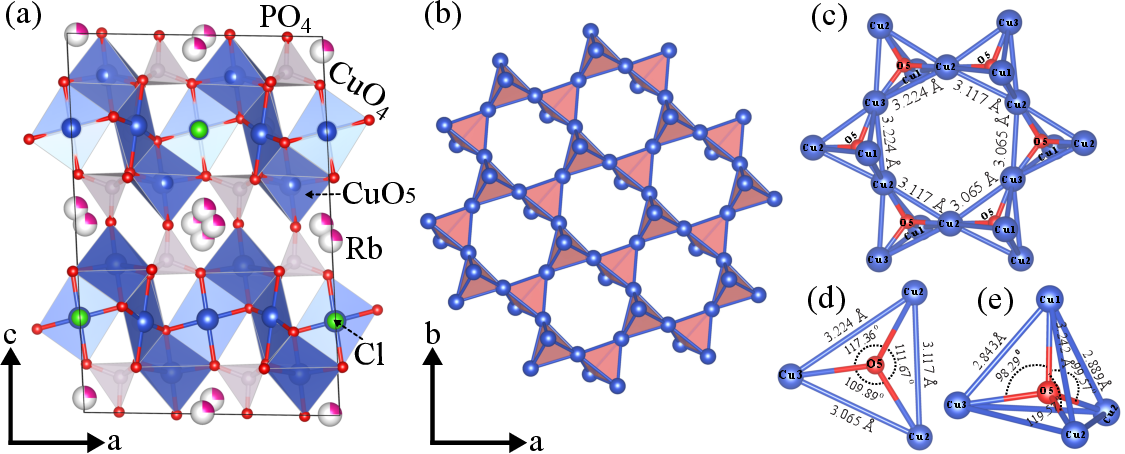} 
\caption{(a) Crystal structure of (RbCl)Cu$_{5}$P$_{2}$O$_{10}$ projected in the $ac$-plane. The connection of Cu(2,3)O$_4$ square planes (plaquettes) and Cu(1)O$_5$ square pyramids forming the capped-kagome layers. PO$_4$ tetrahedra are connecting these layers along the $c$-direction. (b) A section of the capped-kagome layer in the $ab$-plane. (c) A hexagonal ring showing the bond lengths. (d) Triangular base of the OCu$_4$ tetrahedron with bond lengths and bond angles. (e) The bond lengths and bond angles of the apical Cu atom with basal Cu atoms of the OCu$_4$ tetrahedron.}
\label{Fig1}
\end{figure*}
In this paper, we carried out a comprehensive study of the structural and magnetic properties of a phosphate averievite (RbCl)Cu$_{5}$P$_{2}$O$_{10}$, abbreviated as RCCPO. RCCPO is reported to crystallize in a monoclinic structure with space group $C2/c$ (No.~15) at room temperature~\cite{Kornyakov1833}. The crystal structure illustrated in Fig.~\ref{Fig1} consists of capped-kagome layers, made up of corner sharing anion-centered OCu$_4$ tetrahedral units, arranged in an up-down-up-down fashion [see Fig.~\ref{Fig1}(b)].
%\textbf{(These tetrahedral units resemble like pyrochlore slab, which is responsible for the high degree of magnetic frustration in the capped kagome lattice.)}
There are three in-equivalent Cu sites in the formula unit.
Two Cu sites (Cu2 and Cu3) possess distorted CuO$_4$ square planar geometry while the Cu1 site forms a distorted CuO$_5$ square pyramid. The CuO$_4$ squares share corners in order to make the kagome plane and the CuO$_5$ square pyramids share the edges with the adjacent CuO$_4$ squares that form the OCu$_4$ tetrahedra units and hence the capped-kagome network. Here, O5 is the common oxygen that provides the interaction path among two Cu2 and one Cu3 ions at the triangular base and the apical Cu1 ion in an anion-centered OCu$_4$ tetrahedron.
%Two Cu2 and one Cu3 ions form the triangular base while an Cu1 ion capes the OCu$_4$ tetrahedron. and hence, the kagome layers while the Cu1 ion always caps the kagome layers.
%Each anion-centered OCu$_4$ tetrahedra unit shares common corners with three neighbouring tetrahedra, forming hexagonal layers, commonly known as capped-kagome network.
The Cl$^{-}$ ions are located at the center of the hexagonal rings while the Rb$^{+}$ ions are located in the inter-layer spacing.
%The difference between the kagome and capped-kagome lattices is that, a normal kagome lattice consists of corner shared Cu$^{2+}$  triangles whereas a capped-kagome lattice is composed of corner-shared Cu$^{2+}$  triangular-pyramids. 
The non-magnetic PO$_4$ tetrahedra are placed between the capped-kagome layers and connect the capped Cu$^{2+}$ of one layer with Cu$^{2+}$ ions of the neighboring layers. This provides three-dimensional (3D) coupling along the $c$-direction [see Fig.~\ref{Fig1}(a)].
%The apex Cu$^{2+}$ ions in the capped-kagome layers are arranged in a zig-zag fashion above and below the kagome layer as shown in Fig.~\ref{Fig1}(c). Therefore, the capped-kagome lattice is expected to be more frustrated than the conventional kagome lattice.
Further, owing to the low symmetry crystal structure, the Cu$^{2+}$-Cu$^{2+}$ distances within each OCu$_4$ tetrahedral unit are unequal, which induces distortion in the capped-kagome layer [see Figs.~\ref{Fig1}(c - e)].
All these effects render the spin-lattice more complex, opening up the possibility of observing non-trivial ground states. Our magnetic measurements reveal that RCCPO is a highly frustrated magnet and it undergoes two consecutive magnetic transitions at low temperatures.

\section{Experimental Details}
A polycrystalline sample of RCCPO was synthesized by the conventional solid-state reaction technique. The synthesis involves two steps. In the first step, the precursor Cu$_2$P$_2$O$_7$ was prepared by heating the stoichiometric mixture of CuO (Aldrich, 99.999\%) and NH$_2$H$_2$PO$_4$ (Aldrich, 99.999\%). These reagents were finely ground, pressed into pellets, and prereacted in air at 350\,\degree C for 12~h in order to remove ammonia and water. The obtained soft pellets were ground again, sealed in an evacuated quartz tube, and heated at 800\,\degree C for 24~h. In the second step, the obtained Cu$_2$P$_2$O$_7$ powder was mixed in a stoichiometric ratio with CuO (Aldrich, 99.999\%) and RbCl (Aldrich, 99.999\%), ground, pelletized, sealed in an evacuated quartz tube, and annealed at 550 - 570\,\degree C with multiple intermediate re-grindings. The phase purity of the product was confirmed by powder x-ray diffraction (XRD) recorded at room temperature using a PANalytical x-ray diffractometer (Cu\textit{K$_{\alpha}$} radiation, $\lambda_{\rm avg}\simeq 1.5418$ \AA). To check if any structural transition is present, temperature-dependent powder XRD was performed over a broad temperature range (13~K $\leq T \leq 400$~K). For low-temperature measurements, a low-$T$ attachment (Oxford Phenix) and for high-temperature measurements, a high-$T$ oven attachment (Anton-Paar HTK 1200N) to the x-ray diffractometer were used. Rietveld refinement of the powder XRD was performed using the \texttt{FULLPROF} software package~\cite{Carvajal55}, taking the initial structural parameters from the previous report~\cite{Kornyakov1833}.

The dc magnetization ($M$) was measured as a function of temperature (1.9~K~$\leq T \leq 380$~K) in different magnetic fields, using the vibrating sample magnetometer (VSM) attachment to the physical property measurement system [(PPMS) Quantum Design].
%For the high-temperature measurements ($T \geq 380$~K), a high-$T$ oven was attached to the VSM.
Similarly, $ac$ susceptibility was measured as a function of temperature (1.9~K $\leq T \leq 100$~K) and frequency (200~Hz~$\leq \nu \leq 10$~kHz) in an $ac$ field of 10~Oe using the ACMS option of the PPMS. Heat capacity ($C_{\rm p}$) as a function of $T$ and $H$ was measured on a small piece of sintered pellet using the thermal relaxation technique in PPMS.

Nuclear magnetic resonance (NMR) measurements were carried out using a laboratory-built phase-coherent spin-echo pulse spectrometer on the $^{31}$P nuclei (nuclear spin $I=1/2$ and gyromagnetic ratio $\gamma_{N}/2\pi = 17.237$~MHz/T). We perform the experiments at two radio frequencies ($f = 49.6$ and 120.6~MHz) and over a wide temperature range (1.8~K $\leq T \leq$ 300~K). The $^{31}$P NMR spectra were obtained by sweeping the magnetic field, keeping the frequency fixed. The temperature-dependent NMR shift, $K(T)=[H_{\rm ref}-H(T)]/H(T)$ was calculated by taking the resonance field of the sample ($H$) with respect to the resonance field of a nonmagnetic reference ($H_{\rm ref}$) H$_3$PO$_4$. The $^{31}$P spin-lattice relaxation rate ($1/T_1$) was measured by the standard saturation recovery method. $^{31}$P spin-spin relaxation rate ($1/T_2$) was obtained by measuring the decay of the echo integral with variable spacing between the $\pi$/2 and $\pi$ pulses.

\section{Results}
\subsection{X-ray diffraction}
\begin{figure}
	\includegraphics[width=\linewidth]{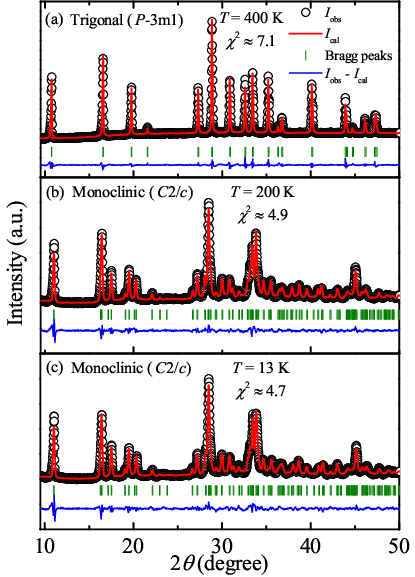}
	\caption{Powder XRD patterns (open circles) at (a) 400~K, (b) 200~K, and (c) 13~K. The solid lines denote the Rietveld refinement of the data. The Bragg peak positions are indicated by green vertical bars and the bottom solid line indicates the difference between the experimental and calculated intensities. The crystal structure and the corresponding space group at different temperatures are also indicated.}
	\label{Fig2} 
\end{figure}
\begin{figure}
	\includegraphics[width=\linewidth]{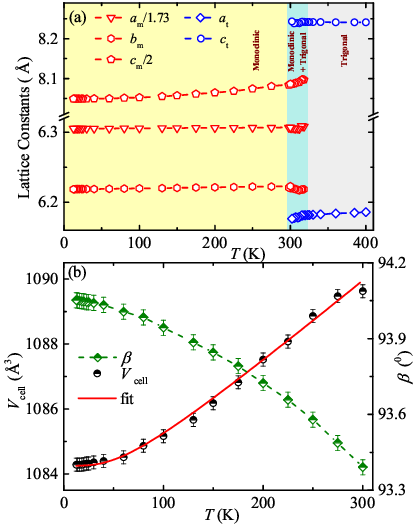}
	\caption{(a) Lattice constants ($a$, $b$, and $c$) as a function of temperature from 13~K to 400~K. The lattice constants in the monoclinic phase are scaled with respect to the high symmetry trigonal structure. The areas shaded in yellow, gray, and cyan colors represent the regimes for pure monoclinic, trigonal, and the coexistence of both phases, respectively. (b) Monoclinic angle (${\beta}$) and the unit cell volume ($V_{\rm cell}$) as a function of temperature from 13~K to 300~K. The solid line represents the fit of $V_{\rm cell}(T)$ using Eq.~\eqref{VcellvsT}.}
	\label{Fig3} 
\end{figure}
In order to detect the structural transition or lattice distortion, if any, powder XRD was measured at different intermediate temperatures from 13~K to 400~K. Figure~\ref{Fig2} displays the powder XRD patterns of RCCPO at 400~K, 200~K, and 13~K along with the Rietveld fit. At room-temperature (300~K), all the diffraction peaks can be well indexed to the monoclinic unit cell with $C2/c$ symmetry. The obtained lattice parameters at room temperature are $a=10.868(4)$~\AA, $b=6.207(5)$~\AA, $c=16.181(5)$~\AA, $\beta=93.391(3)^\circ$, and the unit cell volume $V_{\rm cell}\simeq1089.621$~\AA$^3$, which are in close agreement with the previous report~\cite{Kornyakov1833}. The absence of any visible change in the XRD pattern down to 13~K indicates no structural distortion from 300~K down to 13~K.
%The refined lattice parameters and unit cell volume ($V_{\rm cell}$) are [$a=10.910(4)$~\AA, $b=6.221(5)$~\AA, $c=16.129(5)$~\AA, $\beta=93.724(3)^\circ$, and $V_{\rm cell}\simeq1089.5215$~\AA$^3$] and [$a=10.908(4)$~\AA, $b=6.218(4)$~\AA, $c=16.097(5)$~\AA, $\beta=94.094(3)^\circ$, and $V_{\rm cell}\simeq1084.2861$~\AA$^3$] for $T=200$~K and 13~K, respectively.
However, a change in the diffraction pattern was observed above $T_{\rm t} \sim 310$~K with a splitting of certain Bragg peaks and the appearance of new peaks. Above $\sim 340$~K, the XRD pattern completely stabilizes in a more symmetric crystal structure with less number of Bragg peaks compare to  the room-temperature crystal structure. The analysis of the XRD pattern at 400~K confirmed a trigonal structure with space group $P\bar{3}m1$ (No. 164). The refined lattice parameters are $a=6.185(4)$~\AA, $c=8.241(5)$~\AA, and $V_{\rm cell}\simeq273.1$~\AA$^3$. Thus, a structural transition occurs at around 310~K from a high temperature trigonal to a low temperature monoclinic structure, similar to the analog compound CCCVO. In an intermediate temperature regime around 300~K both the phases co-exist and the data could be fitted using two phases. The obtained atomic parameters at 400~K (trigonal) and 200~K (monoclinic) are listed in Tables~\ref{TableRefinement1} and \ref{TableRefinement2}, respectively.

\begin{table}[ptb]
	\caption{Structural parameters of RCCPO obtained from the Rietveld refinement of the powder-XRD data at 400~K [trigonal, space group: $P\bar{3}m1$ (No. 164)]. Our fit yields $a=6.185(4)$~\AA, $c=8.241(5)$~\AA, $V_{\rm cell}\simeq273.1$~\AA$^3$, $R_{\rm p} \simeq 14.5$, and $\chi^2 = \left(\frac{R_{\rm wp}}{R_{\rm exp}}\right)^2 \simeq 7.2$. Listed are the Wyckoff positions and the refined atomic coordinates for each atom.}
	\label{TableRefinement1}
	\begin{ruledtabular}
		\begin{tabular}{cccccc}
			Atom & Wyckoff & $x$ & $y$ & $z$ & Occ. \\\hline
			Rb1 & $1$a & $1.000$ & $1.000$ & $1.000$ & $0.60$ \\
		    Rb2 & $2$c & $1.000$ & $1.000$ & $0.877$(3) & $0.20$ \\
			Cl1 & $1$b & $1.000$ & $1.000$ & $0.500$ & $1.00$ \\
			Cu1 & $2$d & $0.333$ & $0.666$ & $0.210$(1) & $1.00$\\
			Cu2 & $3$f & $0.000$ & $0.500$ & $0.500$ & $1.00$\\
			P1 & $2$d & $0.333$ & $0.666$ & $0.800$(2) & $1.00$\\
			O1 & $2$d & $0.333$ & $0.666$ & $1.014$(1) & $1.00$\\
			O2 & $6$i & $0.061$(3) & $0.530$(1) & $0.752$(4) & $1.00$\\
			O3 & $2$d & $0.333$ & $0.666$ & $0.452$(1) & $1.00$\\
		\end{tabular}
	\end{ruledtabular}
\end{table}

\begin{table}[ptb]
	\caption{Structural parameters of RCCPO obtained from the Rietveld refinement of the powder-XRD data at 200~K [Monoclinic, Space group: $C2/c$ (No. 15)]. Our fit yields $a=10.860(4)$~\AA, $b=6.221(5)$~\AA, $c=16.130(5)$~\AA, $\beta=93.724(3)^\circ$, $V_{\rm cell}\simeq1087.44$~\AA$^3$, $R_{\rm p} \simeq 14.8$, and $\chi^2 = \left(\frac{R_{\rm wp}}{R_{\rm exp}} \right)^2 \simeq 4.9$. Listed are the Wyckoff positions and the refined atomic coordinates for each atom.}
	\label{TableRefinement2}
	\begin{ruledtabular}
		\begin{tabular}{cccccc}
			Atom & Wyckoff & $x$ & $y$ & $z$ & Occ. \\\hline
			Rb1 & $8$f & $0.506$(2) & $0.028$(2) & $0.546$(2) & $0.25$ \\
			Rb2 & $8$f & $0.471$(2) & $0.033$(2) & $0.506$(2) & $0.25$ \\
			Cl1 & $4$e & $0.500$ & $1.065$(1) & $0.750$ & $1.00$ \\
			Cu1 & $8$f & $0.346$(2) & $0.472$(3) & $0.606$(1) & $1.00$\\
			Cu2 & $8$f & $0.242$(2) & $0.807$(2) & $0.741$(2) & $1.00$\\
			Cu3 & $4$e & $0.500$ & $0.606$(2) & $0.750$ & $1.00$\\
			P1 & $8$f & $0.350$(2) & $0.436$(1) & $0.403$(2) & $1.00$\\
			O1 & $8$f & $0.488$(2) & $0.421$(2) & $0.379$(1) & $1.00$\\
			O2 & $8$f & $0.281$(1) & $0.226$(1) & $0.382$(1) & $1.00$\\
			O3 & $8$f & $0.283$(2) & $0.642$(2) & $0.381$(2) & $1.00$\\
			O4 & $8$f & $0.358$(2) & $0.384$(2) & $0.497$(2) & $1.00$\\
			O5 & $8$f & $0.327$(2) & $0.544$(2) & $0.720$(2) & $1.00$\\
		\end{tabular}
	\end{ruledtabular}
\end{table}
The temperature evolution of lattice parameters from 13~K to 400~K is presented in Fig.~\ref{Fig3}.
%There exists an intermediate temperature regime where both the phases co-exist.
As shown in Fig.~\ref{Fig3}(a), associated with the symmetry lowering, the in-plane and out-of-plane lattice constants transform into $a_{\rm m} \simeq \sqrt{3}a_{\rm t}$, $b_{\rm m} \simeq b_{\rm t}$, and $c_{\rm m} \simeq 2c_{\rm t}$, respectively~\cite{Bader054415,Tsirlin014429}. Here, the subscripts $m$ and $t$ denote the monoclinic and trigonal structures, respectively. Figure~\ref{Fig3}(b) presents the temperature variation of angle ($\beta$) and $V_{\rm cell}$ from 13~K to 300~K in the monoclinic phase. $\beta$ increases with decreasing temperature but $V_{\rm cell}$ is found to decrease systematically upon cooling. The variation of $V_{\rm cell}$ with temperature can be expressed in terms of the internal energy [$U(T)$] of the system~\cite{Guchhait024426}
%\cite{Budd18,Wallace1998}
\begin{equation}
	V_{\rm cell}(T) = \frac{\gamma U(T)}{K_0} + V_0,
	\label{VcellvsT}
\end{equation}
where $V_0$ is the cell volume at $T = 0$~K, $K_0$ is the bulk modulus of the system, and $\gamma$ is the Gr$\ddot{\rm u}$neisen parameter. $U(T)$ is the internal energy which can be expressed in terms of the Debye approximation~\cite{Kittel2004} as
\begin{equation}\label{Uenergy}
	U(T) = 9Nk_{\rm B}T\left(\frac{T}{\theta_{\rm D}}\right)^3 \int_{0}^{\frac{\theta_{\rm D}}{T}}\frac{x^3}{(e^{x}-1)}dx.
\end{equation}
Here, $N$ is the number of atoms in the unit cell, $k_{\rm B}$ is the Boltzmann constant, and $\theta_{\rm D}$ is the Debye temperature. The variable $x$ inside the integration stands for the quantity $\frac{\hbar\omega}{k_{\rm B}T}$ with phonon frequency $\omega$ and Planck constant $\hbar$. Here, $\theta_{\rm D}=\frac{\hbar\omega_{\rm D}}{k_{\rm B}}$ and $\omega_{\rm D}$ is the upper limit of $\omega$. The best fit of the $V_{\rm cell}(T)$ data using Eq.~\eqref{VcellvsT} [solid line in Fig.~\ref{Fig3}(b)] yields the parameters: $\theta_{\rm D}\simeq225$~K, $V_0\simeq$ 1084.25~\AA$^3$, and $\frac{\gamma}{K_0}\simeq8.08\times10^{-12}$~Pa$^{-1}$.

\subsection{Magnetization}
\begin{figure}
	\includegraphics[width=\linewidth]{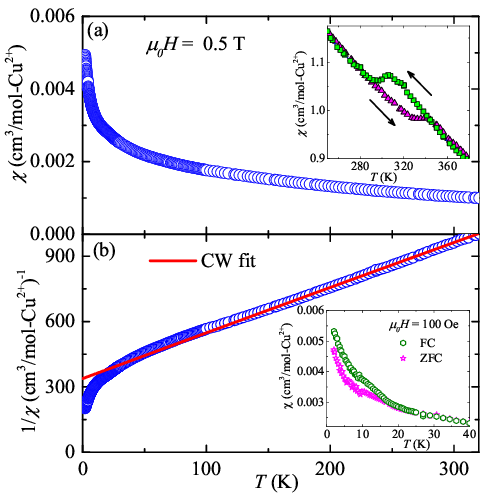}
	\caption{(a) Temperature dependent dc susceptibility $\chi(T)$ measured at an applied field of $\mu_{\rm 0}H = 0.5$~T. Inset: Thermal hysteresis around the structural transition measured under FCW and FCC conditions. (b) Inverse susceptibility ($1/\chi$) vs $T$ at $\mu_{\rm 0}H = 0.5$~T and the solid line is the CW fit for 200~K $\leq T \leq 300$~K below the structural transition. Inset:  $\chi(T)$ measured at $H = 100$~Oe in ZFC and FC protocols.}
	\label{Fig4}
\end{figure}
The temperature dependent dc susceptibility $\chi(T)$ ($\equiv M/H$) of the polycrystalline RCCPO sample measured in the applied field of $\mu_{\rm 0}H = 0.5$~T is shown in the Fig.~\ref{Fig4}(a). At high temperatures, $\chi(T)$ increases with decreasing temperature in a Curie-Weiss (CW) manner as expected in the paramagnetic regime. No clear indication of any magnetic LRO is detected down to 2~K. $\chi(T)$ measured under field-cooled-warming (FCW) and field-cooled-cooling (FCC) conditions shows a clear thermal hysteresis [inset of Fig.~\ref{Fig4}(a)] confirming the structural phase transition.

Figure~\ref{Fig4}(b) shows the inverse magnetic susceptibility $1/\chi(T)$ for $\mu_{\rm 0}H = 0.5$~T. In the paramagnetic regime $1/\chi(T)$ typically shows a linear behavior with temperature, due to uncorrelated moments. To extract the magnetic parameters, $1/\chi(T)$ was fitted in the temperature range 200~K~$\leq T \leq 300$~K below the structural transition by the CW law
\begin{equation}
	\chi(T)=\chi_0+\frac{C}{T-\theta_{\rm CW}},
	\label{cw}
\end{equation}
where, $\chi_0$ is the temperature-independent susceptibility, which includes Van-Vleck paramagnetism and core diamagnetism. The second term is the CW law where $C$ is the Curie constant and $\theta_{\rm CW}$ is the CW temperature. The fit yields $\chi_{0} \simeq 2.19 \times 10^{-4}$~cm$^{3}$/mol-Cu$^{2+}$, $C \simeq 0.46$~cm$^{3}$ K/mol-Cu$^{2+}$, and $\theta_{\rm CW} \simeq -160$~K. The large negative value of $\theta_{\rm CW}$ suggests that the dominant exchange interactions between Cu$^{2+}$ ions are antiferromagnetic (AFM) in nature. From the value of $C$, the effective moment is calculated to be $\mu_{\rm eff} \simeq 1.93\mu_{\rm B}$/Cu$^{2+}$ using the relation $\mu_{\rm eff} = \sqrt{3k_{\rm B}C/N_{\rm A}\mu_{\rm B}^2}$, where $N_{\rm A}$ is the Avogadro's number and $\mu_{\rm B}$ is the Bohr magneton. For a spin-$1/2$ system, the spin-only effective moment is expected to be $\mu_{\rm eff} = g\sqrt{S(S+1)} \mu_{\rm B} \simeq 1.73 \mu_{\rm B}$, assuming a Land$\acute{e}$ $g$-factor $g\simeq 2$. However, our experimental value of $\mu_{\rm eff} \simeq 1.93\mu_{\rm B}$/Cu$^{2+}$ is slightly higher than the spin-only value and corresponds to $g\simeq 2.22$. Such a large value of $g$ is typically observed for powder samples containing magnetic Cu$^{2+}$ ion~\cite{Nath054409}. The core diamagnetic susceptibility $\chi_{\rm core}$ of RCCPO is calculated to be $-2.22 \times 10^{-4}$~cm$^{3}$/mol by adding the core diamagnetic susceptibilities of the individual ions Rb$^{+}$, Cl$^{-}$, Cu$^{2+}$, P$^{5+}$, and O$^{2-}$~\cite{Selwood1956,*Bain532}. The Van-Vleck paramagnetic susceptibility ($\chi_{\rm VV}$) is estimated by subtracting $\chi_{\rm dia}$ from $\chi_{0}$ to be $\sim 4.41 \times 10^{-4}$~cm$^{3}$/mol. This value of $\chi_{\rm VV}$ is close to the values reported for other cuprates~\cite{Guchhait224415,Islam174432}.

As depicted in the inset
of Fig.~\ref{Fig4}(b), the zero-field-cooled (ZFC) and field-cooled (FC) susceptibilities in $H = 100$~Oe show a weak splitting at $T_{\rm N1} \simeq 20$~K, indicating either the onset of a magnetic LRO or a spin-glass type transition. The extent of frustration in a spin system can be quantified
by the frustration ratio $f = \frac{|\theta_{\rm CW}|}{T_{\rm N1}}$. According to the mean field theory, $\theta_{\rm CW}$ is the sum of all exchange interactions present in the system. Typically, for a non-frustrated AFM system, $f$ is close to 1. However, for a highly frustrated AFM, the $f$ value is much larger than 1. For RCCPO, the frustration ratio is calculated to be $f \simeq 160/20 \simeq 8$ which corroborates the strong frustration in the system.

\begin{figure}
	\includegraphics[width=\linewidth]{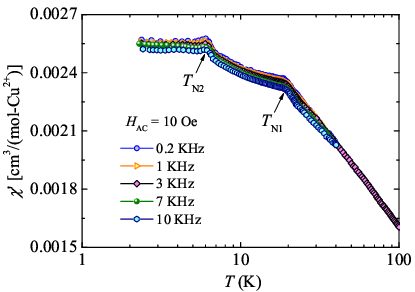}
	\caption{Real part of ac susceptibility $\chi'(T)$ vs $T$ at different frequencies. The arrows point to the magnetic anomalies.}
	\label{Fig5}
\end{figure}
Figure~\ref{Fig5} presents the temperature dependent ac susceptibility measured in different frequencies and at a fixed ac field of $H_{\rm AC} \simeq 10$~Oe. In contrast to dc $\chi(T)$, the real part of the ac susceptibility $\chi'(T)$ shows two anomalies at $T_{\rm N1} \simeq 20$~K and $T_{\rm N2} \simeq 7$~K, reflecting two magnetic transitions. The peak at $T_{\rm N1} \simeq 20$~K is found to be weakly frequency dependent, which shifts towards higher temperatures with increasing frequency. This is a possible indication of canted antiferromagnetism or a spin-glass (SG) transition.  On the other hand, the peak at $T_{\rm N2} \simeq 7$~K is found to be frequency independent, suggesting the onset of a robust AFM LRO.
 
\subsection{Heat capacity}
\begin{figure}
	\includegraphics[width=\linewidth]{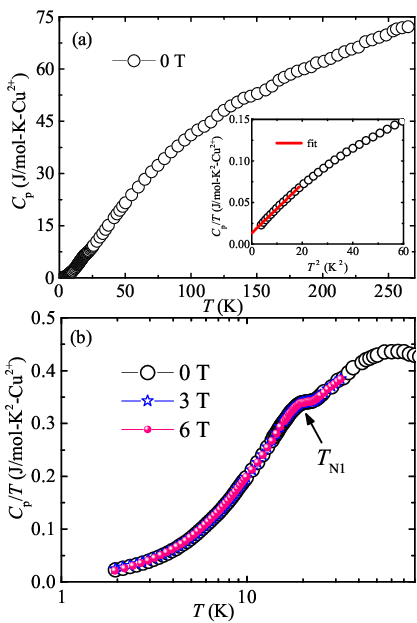}
	\caption{(a) $C_{\rm p}(T)$ of RCCPO measured in zero-field. Inset: $C_{\rm p}/T$ vs $T^{2}$ in zero-field, showing the linear regime below $T_{\rm N2}$. Solid line is the linear fit. (b) $C_{\rm p}/T$ vs $T$ measured in different applied fields in the low-temperature region, around $T_{\rm N1}$.}
	\label{Fig6}
\end{figure}
Temperature dependent heat capacity $C_{\rm p}(T)$ measured in zero-field is presented in Fig.~\ref{Fig6}(a). At high temperatures, $C_{\rm p}$ is dominated by the phonon excitations while at low temperatures, it is dominated by the magnetic contribution. As the temperature is lowered, $C_{\rm p}$ shows a weak cusp at $T_{\rm N1} \simeq 20$~K [see Fig.~\ref{Fig6}(b)]. The position of this cusp is unaffected by the external magnetic field up to 6~T. However, no obvious feature is evident at $T_{\rm N2} \simeq 7$~K, likely due to a weak entropy change across $T_{\rm N2}$. Below 4~K, the spin-wave dispersion gives rise to a $T^3$ behavior, as expected in a 3D AFM ordered state~\cite{Islam174432}. The plot of $C_{\rm p}/T$ vs $T^{2}$ in the inset of Fig.~\ref{Fig6}(a) highlights the linear behavior below 4~K. Moreover, in case of a SG transition, $C_{\rm p}(T)$ at low temperatures deviates from the $T^3$ behavior and is usually described by $C_{\rm p}(T) = \beta T^3 + \delta T^{3/2}$~\cite{Islam134433}. Thus, only the $T^3$ dependence of $C_{\rm p}(T)$ also rules out a SG transition at $T_{\rm N2}$ and establishes the canted-AFM nature of the transition.
%The value of $C_{\rm p}$ at $T = 250$~K is about 75.2 J/(mol-K-Cu$^{2+}$) which is equal to 376 J/(mol-K). This is lesser than the expected Dulong-Petit value $C_{\rm v} = 3mR = 57R = 473.8$ J/(mol-K), where $R$ is the the gas constant and $m$ is the number of atoms per formula unit.
Due to the unavailability of a suitable non-magnetic analog, we are unable to separate the magnetic part of the heat capacity from the total $C_{\rm p}(T)$.

\subsection{$^{31}$P NMR}
NMR is an immensely powerful local tool to study the static and dynamic properties of a spin system. The crystal structure of RCCPO has a unique $^{31}$P site. Two adjacent capped-kagome layers in the $ac$-plane are connected through the PO$_4$ tetrahedra. Since $^{31}$P is coupled with the Cu$^{2+}$ ions, through $^{31}$P NMR, one can probe the static and dynamic properties of Cu$^{2+}$ spins.

\subsubsection{$^{31}$P NMR spectra}
\begin{figure}
	\includegraphics[width=\linewidth]{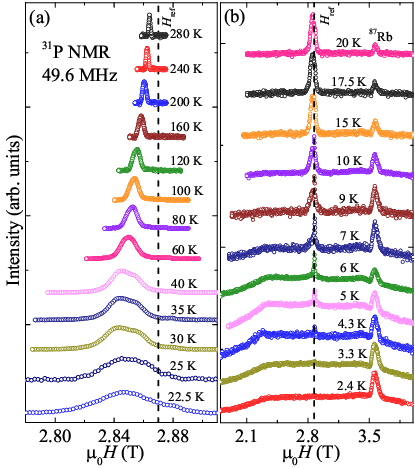}
	\caption{(a) Temperature evolution of $^{31}$P NMR spectra measured at a radio frequency $f = 49.6$~MHz, above $T_{\rm N1}$. The vertical dashed line corresponds to the $^{31}$P non-magnetic reference field position. (b) $^{31}$P NMR spectra below $T_{\rm N1}$. The peak at the right side corresponds to the signal of $^{87}$Rb present in the sample.}
	\label{Fig7}
\end{figure}
The field-sweep $^{31}$P NMR spectra measured at different temperatures (2.4~K$\leq T\leq 280$~K) in a radio frequency of 49.6~MHz are shown in Fig.~\ref{Fig7}. Each NMR spectrum is normalized by its maximum amplitude and offset vertically by adding a constant. The spectral line consists of a single spectral line, typical for $I=1/2$ nuclei~\cite{Mukharjee144433}. The spectral line is symmetric at high temperatures and becomes asymmetric as we go down in temperature. Since our measurements are done on a randomly oriented polycrystalline sample, the asymmetric shape of the spectra can be attributed to a powder pattern due to an asymmetric hyperfine coupling constant and/or an anisotropic susceptibility~\cite{Yogi024413}. With decreasing temperature, the line broadens and the peak position shifts with temperature.

\begin{figure}
	\includegraphics[width=\linewidth]{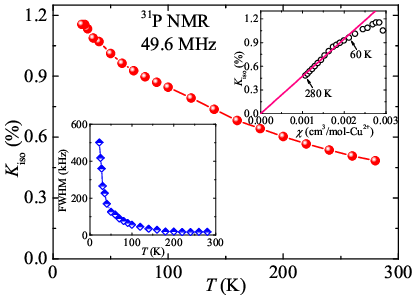}
	\caption{Temperature dependent $^{31}$P NMR iso-shift $K_{\rm iso}$ measured at 49.6~MHz. Upper inset: $K_{\rm iso}$ vs $\chi$ (measured at 3~T). The solid line is the straight line fit. Lower inset: Full width at half maximum (FWHM) vs $T$.}
	\label{Fig8}
\end{figure}
The NMR shift $K(T)$ is the direct measure of intrinsic spin susceptibility $\chi_{\rm spin}(T)$ and is free from extrinsic contributions. Therefore, one can write $K(T)$ in terms of $\chi_{\rm spin}(T)$ as
\begin{equation}
	\label{K}
	K(T) = K_0 +\frac{A_{\rm hf }}{N_{\rm A}}\chi_{\rm spin}(T),
\end{equation}
where, $K_0$ is the temperature-independent chemical shift and $A_{\rm hf}$ is the hyperfine coupling constant between the $^{31}$P nuclei and Cu$^{2+}$ electronic spins. The isotropic NMR shift $K_{\rm iso}$ obtained by fitting the NMR spectra is plotted as a function of $T$ in Fig.~\ref{Fig8}.
The upper inset of Fig.~\ref{Fig8} shows the $K_{\rm iso}$ vs $\chi$ plot with $T$ as an implicit parameter. A straight line fit over a temperature range 60~K$\leq T\leq$ 280~K yields the isotropic part of the hyperfine coupling $A_{\rm iso} \simeq 2.93$~T/$\mu_{\rm B}$. The small deviation from linearity below 60~K can be attributed to a small amount of paramagnetic impurities to which $K_{\rm iso}$ is insensitive. The full width at half maximum (FWHM) obtained from the spectral fit is plotted against $T$ in the lower inset of Fig.~\ref{Fig8}. It increases with lowering temperature and then shoots up below about 30~K, which suggests the growth of the internal field as we approach the magnetic ordering at $T_{\rm N1}$.

Figure~\ref{Fig7}(b) presents $^{31}$P NMR spectra below $T_{\rm N1}$. Below about 22~K, the NMR line broadens drastically, implying that $^{31}$P senses the internal field in the ordered state. The spectrum seems to form a triangular line shape, typically expected for a powder sample in an incommensurate spin-density-wave (SDW) state~\cite{Kontani672,Sakurai024428,Ranjith014415,Ranjith024422}. 
Below the second transition $T_{\rm N2} \simeq 9.5$~K,
a huge internal field pops up leading to a drastic line broadening, and the line attains a nearly
rectangular shape superimposed with a sharp line at the center of gravity $H \simeq 2.87$~T, which
is the zero-shift resonance position of $^{31}$P nuclei. With decreasing temperature, this line broadening increases and the intensity of the central narrow line decreases. This rectangular line shape is reminiscent of a commensurate AFM ordering~\cite{Ranjith014415,Ranjith024422,Yamada1751,Kikuchi2660}. An additional signal appears at low temperatures which overlaps with the right shoulder of the rectangular pattern. This peak position is almost temperature independent and corresponds to the signal of $^{87}$Rb present in the sample.

\subsubsection{Spin-lattice relaxation rate $1/T_1$}
\begin{figure}
	\includegraphics[width=\linewidth]{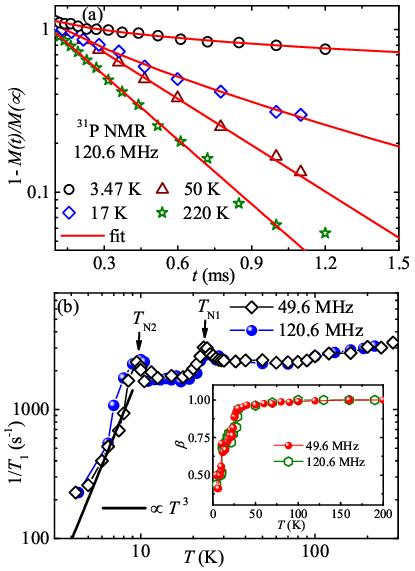}
	\caption{(a) Longitudinal magnetization recovery curves for 120.6~MHz at four selective temperatures measured on the $^{31}$P nuclei and the solid lines are fits using Eq.~\eqref{exp}. (b) $^{31}$P NMR spin-lattice relaxation rate ($1/T_{1}$) vs $T$ measured at 49.6~MHz and 120.6~MHz. The downward arrows point to $T_{\rm N1}\simeq 23.5$~K and $T_{\rm N2}\simeq 9.5$~K. The solid line represents $T^3$ behavior below $T_{\rm N2}$. Inset: Exponent $\beta$ as a function of $T$ for both the frequencies.}
	\label{Fig9}
\end{figure}
To understand the local spin-spin correlation, the $^{31}$P spin-lattice relaxation rate $1/T_1$ was measured as a function of temperature down to 2.4~K at the central peak position and at two different frequencies, 49.6~MHz and 120.6~MHz. For an $I = 1/2$ nucleus, the recovery of the longitudinal magnetization is expected to follow a single exponential behavior. Indeed, our recovery curves were fitted well by the stretched exponential function
\begin{equation}
	1-\frac{M(t)}{M(\infty)}= Ae^{-(t/T_{1})^\beta},
	\label{exp}
\end{equation}
where, $M(t)$ is the nuclear magnetization at a time $t$ after the saturation pulse, $M(\infty)$ is the equilibrium nuclear magnetization, and $\beta$ is the stretch exponent. Recovery curves for 120.6~MHz in four different temperatures along with the fits are shown in Fig.~\ref{Fig9}(a).

The temperature dependence of $1/T_1$ extracted following the above fitting procedure is presented in Fig.~\ref{Fig9}(b) for both the frequencies. The exponent $\beta$ as a function of $T$ for both the frequencies is also plotted in the inset of Fig.~\ref{Fig9}(b). At high temperatures ($T > 30$~K), the value of $\beta$ is found to be close to 1 and is almost temperature independent, suggesting a uniform relaxation process in this temperature range. However, a drastic drop in the $\beta$ value below about 25~K indicates the distribution of relaxation time in the ordered state which is possibly due to some kind of disorder at the magnetic sites~\cite{Johnston176408}. Similarly, $1/T_1$ in the high temperature regime ($T>75$~K) is almost temperature independent due to localized moments in the paramagnetic state~\cite{Moriya23}. At low-temperatures, $1/T_{1}$ exhibits two sharp peaks at around $T_{\rm N1}\simeq 23.5$~K and $T_{\rm N2}\simeq 9.5$~K, indicating the slowing down of fluctuating moments as we approach the magnetic transitions. These findings corroborate the double transitions observed from the $ac$ susceptibility measurements. Note that the transition anomalies in $1/T_1$ appear at slightly higher temperatures compared to that observed in $ac$ susceptibility data, which is due to different thermocouples with different calibrations used in the NMR and PPMS cryostats. Below $T_{\rm N2}$, $1/T_{1}$ drops swiftly toward zero because of the release of critical fluctuations and the scattering of magnons by the nuclear spins~\cite{Beeman359,Belesi184408,Nath214430}.
%In the magnetically ordered state ($T<T_{\rm N}$), the strong temperature dependence of $1/T_1$ is a clear signature of the relaxation due to scattering of magnons by the nuclear spins~\cite{Belesi184408}.
For $T\gg\Delta/k_{\rm B}$, $1/T_1$ follows either a $T^3$ behavior or a $T^5$ behavior due to a two-magnon Raman process or a three-magnon process, respectively, where $\Delta/k_{\rm B}$ is the energy gap in the spin-wave excitation spectrum~\cite{Beeman359}. On the other hand, for $T\ll\Delta/k_{\rm B}$, it follows an activated behavior $1/T_1 \propto T^2e^{-\Delta/k_{\rm B}T}$. As shown in Fig.~\ref{Fig9}(b), $1/T_1$ below $T_{\rm {N2}}$ follows a nearly $T^3$ behaviour, ascertaining that the relaxation is mainly governed by a two magnon process similar to that reported for other frustrated magnets~\cite{Nath214430}.

%At the 3D ordering temperature, the correlation length is expected to diverge, and $1/T_1$ in a narrow temperature range just above $T_{\rm N}$ (i.e., in the critical regime) should be described by the powder law, $1/T_1 \propto \tau^{-\gamma}$, where $\gamma$ is the critical exponent. The value of $\gamma$ represents universal property of the spin system depending upon its dimensionality, the symmetry of the spin lattice, and the type of interactions. To analyze this critical behaviour, $1/T_1$ is plotted against the reduced temperature $\tau$. The data just above $T_{\rm N1}$($\tau \leq 0.4$) were fitted by the powder law with a fixed $T_{\rm N1}\simeq 24$~K yielding $\gamma \simeq 0.11$. For the 3D Heisenberg antiferromagnet, a mean-field theory predicts $\gamma = 1/2$ and a dynamic scaling theory gives $\gamma = 1/3$~\cite{Lee214416}. Our experimental value $\gamma \simeq 0.11$ is far below these theoretically predicted values. This effect requires further investigation and may be intertwined with the strong frustration in 3D that affects spin dynamics above $T_{\rm N1}$.   

From the constant value of $1/T_1$ at high-temperatures, one can estimate the leading exchange coupling between Cu$^{2+}$ ions, using the hyperfine coupling between P and Cu atoms. At high temperatures, $1/T_1$ can be expressed as~\cite{Moriya23,Nath214430}
\begin{equation}
\left(\frac{1}{T_1}\right)_{T\rightarrow\infty} =
\frac{(\gamma_{N} g\mu_{\rm B})^{2}\sqrt{2\pi}z^\prime S(S+1)}{3\,\omega_{ex}}
\left(\frac{A_{z}}{z^\prime}\right)^{2},
\label{t1inf}
\end{equation}
where $\omega_{ex}=\left(|J^{\rm max}|k_{\rm B}/\hbar\right)\sqrt{2zS(S+1)/3}$ is the Heisenberg exchange frequency, $z$ is the number of nearest-neighbor spins of each Cu$^{2+}$ ion, and $z^\prime$ is the number of nearest-neighbor Cu$^{2+}$ spins attached to a given P site. In RCCPO, each Cu$^{2+}$ ion in the kagome plane has six nearest-neighbors while each capped Cu$^{2+}$ ion can see only three neighboring ions. Therefore, on an average each Cu$^{2+}$ spin sees 4.5 neighboring spins. Similarly, each P site is strongly connected to five nearest-neighbour Cu$^{2+}$ spins. Thus, using the parameters $A_{\rm z} \simeq 2.93$~T/$\mu_{\rm B}$, $\gamma_N = 108.303 \times 10^2\,{\rm rad}$~sec$^{-1}$\,Oe$^{-1}$, $z=4.5$, $z^\prime=5$, $g=2.22$, $S=\frac12$, and the relaxation rate at 160~K $\left(\frac{1}{T_1}\right)_{T\rightarrow\infty}\simeq 2714.7$ sec$^{-1}$, the magnitude of the leading antiferromagnetic exchange coupling is calculated to be $J^{\rm max}/k_{\rm B}\simeq 117$~K.

\subsubsection{Spin-spin relaxation rate $1/T_2$}
\begin{figure}
	\includegraphics[width=\linewidth]{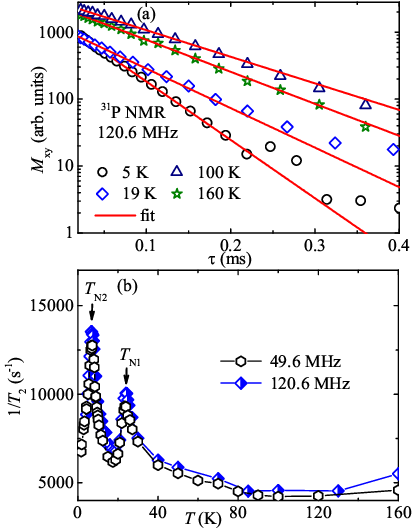}
	\caption{(a) Transverse magnetization recovery curves as a function of $\tau$ at four different temperatures. The solid lines show the fit using Eq.~\eqref{exp2}. (b) $^{31}$P NMR spin-spin relaxation rate ($1/T_2$) vs $T$ measured at 49.6~MHz and 120.6~MHz. The downward arrows mark the transition temperatures at $T_{\rm N1}\simeq 23.5$~K and $T_{\rm N2}\simeq 9.5$~K.}
	\label{Fig10}
\end{figure}
In order to measure the spin-spin relaxation rate $1/T_2$, the decay of the transverse magnetization
($M_{\rm xy}$) was monitored after a $\pi/2$ - $\tau$ - $\pi$ pulse sequence as a function of the pulse separation time $\tau$. The recovery curves are then fitted by the following equation
\begin{equation}
	M_{\rm xy}= M_0 e^{-(2\tau/T_{2})}.
	\label{exp2}
\end{equation}
Recovery curves at a few selected temperatures along with the fits are depicted in Fig.~\ref{Fig10}(a). The extracted $1/T_2$ is plotted as a function of temperature in Fig.~\ref{Fig10}(b). Similar to $1/T_1$, $1/T_2$ also exhibits two sharp peaks, further establishing double transitions at $T_{\rm N1}\simeq 23.5$~K and $T_{\rm N2}\simeq 9.5$~K.

\section{Discussion and Summary}
%\textbf{RCCPO is found to be different from its previously known vanadate analogue CCCVO. The major factors responsible for this difference is "chemical pressure" resulting from a much smaller size of PO$_4$ unit, compared to VO$_4$ unit and substitution of Rb in place of Cs. This substitution leads to significant changes in the arrangement of oxygen atoms around the Cu1 site that change its coordination from triangular pyramidal to square pyramidal.}

This averievite family of compounds is an interesting class of compounds where the replacement of any ion in the structure affects the geometry of the OCu$_4$ tetrahedron and alters the magnetic properties significantly. %alkali metal cation and  in the structures of the averievite-type phosphates also affects the geometry of OCu4 tetrahedra.
For instance, in the phosphates (CsCl,CsBr,CsI)Cu$_5$P$_2$O$_{10}$, the value of $\theta_{\rm CW}$ which represents the overall energy scale of the exchange couplings as well as the value of $T_{\rm N}$ increase systematically as the halide ion is changed from Cl to I in the order of increasing ionic radius~\cite{Winiarski4328}.
%Usually, the nature and strength of exchange interaction depends on the Cu - Cu bond distance and the Cu-O-Cu bond angle. As one replaces the halide ion by Cl, Br, and I, it is found that the angle within the kagome plane increases while the angle made by the capped Cu$^{2+}$ with the Cu$^{2+}$ ions in the kagome plane decreases systematically. All the Cu-Cu distances are also increasing systematically as the halide ion is changed from Cl to I. This possibly result in a stronger AFM coupling within the kagome plane but weaker interaction with the capping layer, leading to a reduction in overall exchange coupling.
%\textbf{Similarly presence of Rb$^{+}$ ion in RCCPO in place of Cs$^{+}$ ion in (CsCl)Cu$_5$V$_2$O$_{10}$ and (CsCl)Cu$_5$P$_2$O$_{10}$, contract the interlayer distance because of its smaller ionic radius, which may increase the 3D coupling along $c$-direction. Also, the substitution of Cs by Rb leads to significant changes in the arrangement of oxygen atoms around the Cu1 site that change its coordination from triangular pyramidal of other compounds in this series to square pyramidal in RCCPO.}
The obtained values of $\theta_{\rm CW}$ and $T_{\rm N}$ for RCCPO are comparable to that of the analog compound CCCVO, resulting in an almost identical frustration parameter ($f \simeq 8$ for RCCPO, $f \simeq 7.7$ for CCCVO). A recent theoretical work has shown that upon the replacement of V$^{5+}$ by P$^{5+}$ in CCCVO where the ionic radius of P$^{5+}$ is two times smaller than that of V$^{5+}$, the coupling within the kagome layer remains almost the same whereas the coupling between a kagome layer and capped Cu atom becomes five times larger, increasing the degree of frustration within the OCu$_4$ tetrahedra~\cite{Dey125106}. In this analogy, RCCPO should have a larger $\theta_{\rm CW}$ value than CCCVO because of the replacement of V$^{5+}$ by P$^{5+}$. Here, another difference is that RCCPO contains Rb$^{+}$ that has a smaller ionic radius than Cs$^{+}$ in CCCVO, which possibly plays a crucial role due to which the overall energy scale of the exchange couplings is comparable.

In contrast to a single magnetic transition in all phosphate compounds [(CsCl,CsBr,CsI)Cu$_5$P$_2$O$_{10}$], there appears two magnetic transitions ($T_{\rm N1} \simeq 20$~K and $T_{\rm N2} \simeq 7$~K) in RCCPO at low temperatures which makes this compound an exception in this series. It is to be noted that, despite having the same crystal structure (monoclinic), their space group symmetry is found to be different ($C2/c$ for RCCPO and $P2_{1}/c$ for the other phosphates). The OCu$_4$ tetrahedra in RCCPO are highly distorted compared to the other phosphates. In RCCPO, each tetrahedron consists of three in-equivalent Cu sites in contrast to two Cu sites in other phosphates. A diligent scrutiny of the crystal structure [see Figs.~\ref{Fig1}(c) - (e)] reveals that Cu-Cu bond distances and $\angle Cu-O-Cu$ bond angles in the triangular base of the tetrahedron are inequivalent for RCCPO in contrast to the equilateral triangular base in other phosphates. Similarly, the Cu-Cu bond distances and $\angle Cu-O-Cu$ bond angles of the apical (or capped) Cu with respect to the Cu atoms at the base are all in-equivalent for RCCPO in contrast to other phosphates. Moreover, the six Cu-Cu bond distances in the hexagonal ring [Fig.~\ref{Fig1}(c)] are also found to be unequal and vary from 3.065 to 3.225~\AA for RCCPO while this hexagon is isotropic with the same bond distance for all other phosphates. Thus, the strong distortion in the crystal structure could be the reason for having different ground state properties in RCCPO.

Double magnetic transitions are predicted to occur in anisotropic triangular lattice antiferromagnets (TLAFs) when the magnetic anisotropy is of easy-axis type~\cite{Matsubara2424,Miyashita3385,Quirion014414,Ranjith014415}.
%while the TLAFs with easy-plane type anisotropy may show only single transition~\cite{Matsubara2424,Miyashita3385,Quirion014414,Ranjith014415}.
On lowering the temperature, the collinear up-up-down state appears before the $120\degree$ state and the temperature range of the intermediate phase ($T_{\rm N1}-T_{\rm N2})/T_{\rm N1}$ reflects the relative strength of the easy-axis anisotropy with respect to the isotropic intralayer coupling. Experimentally, double transitions are reported in several TLAFs~\cite{Lee104420,Zhou267206,Ranjith115804,Yokota014403,Zhou267206,Yokota014403}.
%Thus, the two consecutive magnetic transitions at low temperatures is attributed to the easy-axis anisotropy in our compounds and the $120\degree$ ordering below $T_{\rm N2}$ is preceded by a collinear order between $T_{\rm N1}$ and $T_{\rm N2}$. The observed narrow temperature regime between $T_{\rm N1}$ and $T_{\rm N2}$ in both the compounds suggests considerably weak anisotropy compared to the isotropic exchange coupling in the Hamiltonian~\cite{Shirata057205}. %This implies that these compounds are more close to Heisenberg model in contrast to strong anisotropy anticipated for Co-based compounds~\cite{Line546,*Shiba2326} and the local environment of Co$^{2+}$ is close to a cubic symmetry as in KCoF$_3$ and Ba$_3$CoSb$_2$O$_9$~\cite{Shirata057205}.
Double magnetic transitions in zero-field are also observed in staircase-kagome magnets (Mn,Ni,Co)$_3$V$_2$O$_8$ and PbCu$_3$TeO$_7$~\cite{Morosan144403,Yoo2397,Rai235101,Wilson094432}. In (Mn,Ni,Co)$_3$V$_2$O$_8$, both the phases are reported to be incommensurate in nature, originating from
two decoupled sub-lattices in the kagome staircase
structure~\cite{Dong024427}. However, in RCCPO, the two transitions are different. Our $^{31}$P NMR spectral measurements elucidate probably an incommensurate-type phase below $T_{\rm N1}$ and a commensurate phase below $T_{\rm N2}$. Since the capped-kagome structure is a corner sharing of triangular motifs, the two successive transitions observed can be possibly attributed to anisotropy in the compound. However, a more precise knowledge about the nature and origin of the transitions can only be obtained from neutron scatting experiments and complementary theoretical calculations.

%Two-step magnetic transitions are even reported recently in one-dimensional chain compound Pb$_3$TeCo$_3$V$_2$O$_{14}$, likely due to magnetic anisotropy~\cite{Markina104409}.

In summary, we present the structural and magnetic properties of a geometrically frustrated quantum magnet (RbCl)Cu$_{5}$P$_{2}$O$_{10}$. Below $T_{\rm t} \simeq 310$~K, the monoclinic structure portrays a capped kagome lattice of Cu$^{2+}$ ions. Sizable magnetic frustration is gauged from a large value of the frustration parameter. The dual magnetic transitions at low temperatures where a commensurate AFM ordering is preceded by an incommensurate ordering in a capped kagome lattice is unusual and can be possibly ascribed to strong distortion and/or magnetic anisotropy in the spin-lattice. Nevertheless, these ambiguous features call for further experimental investigations on this compound.

\acknowledgments
We would like to acknowledge SERB, India for financial support bearing sanction Grant No.~CRG/2022/000997. Work at the Ames National Laboratory was supported by the U.S. Department of Energy, Office of Science, Basic Energy Sciences, Materials Sciences and Engineering Division. The Ames National Laboratory is operated for the U.S. Department of Energy by Iowa State University under Contract No.~DEAC02-07CH11358.

%\bibliography{ref_RCCPO}

\begin{thebibliography}{63}%
	\makeatletter
	\providecommand \@ifxundefined [1]{%
		\@ifx{#1\undefined}
	}%
	\providecommand \@ifnum [1]{%
		\ifnum #1\expandafter \@firstoftwo
		\else \expandafter \@secondoftwo
		\fi
	}%
	\providecommand \@ifx [1]{%
		\ifx #1\expandafter \@firstoftwo
		\else \expandafter \@secondoftwo
		\fi
	}%
	\providecommand \natexlab [1]{#1}%
	\providecommand \enquote  [1]{``#1''}%
	\providecommand \bibnamefont  [1]{#1}%
	\providecommand \bibfnamefont [1]{#1}%
	\providecommand \citenamefont [1]{#1}%
	\providecommand \href@noop [0]{\@secondoftwo}%
	\providecommand \href [0]{\begingroup \@sanitize@url \@href}%
	\providecommand \@href[1]{\@@startlink{#1}\@@href}%
	\providecommand \@@href[1]{\endgroup#1\@@endlink}%
	\providecommand \@sanitize@url [0]{\catcode `\\12\catcode `\$12\catcode
		`\&12\catcode `\#12\catcode `\^12\catcode `\_12\catcode `\%12\relax}%
	\providecommand \@@startlink[1]{}%
	\providecommand \@@endlink[0]{}%
	\providecommand \url  [0]{\begingroup\@sanitize@url \@url }%
	\providecommand \@url [1]{\endgroup\@href {#1}{\urlprefix }}%
	\providecommand \urlprefix  [0]{URL }%
	\providecommand \Eprint [0]{\href }%
	\providecommand \doibase [0]{https://doi.org/}%
	\providecommand \selectlanguage [0]{\@gobble}%
	\providecommand \bibinfo  [0]{\@secondoftwo}%
	\providecommand \bibfield  [0]{\@secondoftwo}%
	\providecommand \translation [1]{[#1]}%
	\providecommand \BibitemOpen [0]{}%
	\providecommand \bibitemStop [0]{}%
	\providecommand \bibitemNoStop [0]{.\EOS\space}%
	\providecommand \EOS [0]{\spacefactor3000\relax}%
	\providecommand \BibitemShut  [1]{\csname bibitem#1\endcsname}%
	\let\auto@bib@innerbib\@empty
	%</preamble>
	\bibitem [{\citenamefont {Ramirez}(1994)}]{Ramirez453}%
	\BibitemOpen
	\bibfield  {author} {\bibinfo {author} {\bibfnamefont {A.~P.}\ \bibnamefont
			{Ramirez}},\ }\bibfield  {title} {\bibinfo {title} {{Strongly geometrically
				frustrated magnets}},\ }\href
	{https://doi.org/10.1146/annurev.ms.24.080194.002321} {\bibfield  {journal}
		{\bibinfo  {journal} {Annu. Rev. Mater. Sci.}\ }\textbf {\bibinfo {volume}
			{24}},\ \bibinfo {pages} {453} (\bibinfo {year} {1994})}\BibitemShut
	{NoStop}%
	\bibitem [{\citenamefont {Diep}()}]{Diep2013}%
	\BibitemOpen
	\bibfield  {author} {\bibinfo {author} {\bibfnamefont {H.~T.}\ \bibnamefont
			{Diep}},\ }\href@noop {} {\emph {\bibinfo {title} {{Frustrated Spin
					Systems}}}}\ (\bibinfo  {publisher} {World scientific)(2013})\BibitemShut
	{NoStop}%
	\bibitem [{\citenamefont {Balents}(2010)}]{Balents199}%
	\BibitemOpen
	\bibfield  {author} {\bibinfo {author} {\bibfnamefont {L.}~\bibnamefont
			{Balents}},\ }\bibfield  {title} {\bibinfo {title} {{Spin liquids in
				frustrated magnets}},\ }\href {https://doi.org/10.1038/nature08917}
	{\bibfield  {journal} {\bibinfo  {journal} {Nature}\ }\textbf {\bibinfo
			{volume} {464}},\ \bibinfo {pages} {199} (\bibinfo {year}
		{2010})}\BibitemShut {NoStop}%
	\bibitem [{\citenamefont {Yan}\ \emph {et~al.}(2011)\citenamefont {Yan},
		\citenamefont {Huse},\ and\ \citenamefont {White}}]{Yan1173}%
	\BibitemOpen
	\bibfield  {author} {\bibinfo {author} {\bibfnamefont {S.}~\bibnamefont
			{Yan}}, \bibinfo {author} {\bibfnamefont {D.~A.}\ \bibnamefont {Huse}},\ and\
		\bibinfo {author} {\bibfnamefont {S.~R.}\ \bibnamefont {White}},\ }\bibfield
	{title} {\bibinfo {title} {{Spin-Liquid Ground State of the
				spin-$\frac{1}{2}$ Kagome Heisenberg Antiferromagnet}},\ }\href
	{https://doi.org/10.1126/science.1201080} {\bibfield  {journal} {\bibinfo
			{journal} {Science}\ }\textbf {\bibinfo {volume} {332}},\ \bibinfo {pages}
		{1173} (\bibinfo {year} {2011})}\BibitemShut {NoStop}%
	\bibitem [{\citenamefont {Han}\ \emph {et~al.}(2012)\citenamefont {Han},
		\citenamefont {Helton}, \citenamefont {Chu}, \citenamefont {Nocera},
		\citenamefont {Rodriguez-Rivera}, \citenamefont {Broholm},\ and\
		\citenamefont {Lee}}]{Han406}%
	\BibitemOpen
	\bibfield  {author} {\bibinfo {author} {\bibfnamefont {T.~H.}\ \bibnamefont
			{Han}}, \bibinfo {author} {\bibfnamefont {J.~S.}\ \bibnamefont {Helton}},
		\bibinfo {author} {\bibfnamefont {S.}~\bibnamefont {Chu}}, \bibinfo {author}
		{\bibfnamefont {D.~G.}\ \bibnamefont {Nocera}}, \bibinfo {author}
		{\bibfnamefont {J.~A.}\ \bibnamefont {Rodriguez-Rivera}}, \bibinfo {author}
		{\bibfnamefont {C.}~\bibnamefont {Broholm}},\ and\ \bibinfo {author}
		{\bibfnamefont {Y.~S.}\ \bibnamefont {Lee}},\ }\bibfield  {title} {\bibinfo
		{title} {{Fractionalized excitations in the spin-liquid state of a
				kagome-lattice antiferromagnet}},\ }\href
	{https://doi.org/10.1038/nature11659} {\bibfield  {journal} {\bibinfo
			{journal} {Nature}\ }\textbf {\bibinfo {volume} {492}},\ \bibinfo {pages}
		{406} (\bibinfo {year} {2012})}\BibitemShut {NoStop}%
	\bibitem [{\citenamefont {Khuntia}\ \emph {et~al.}(2020)\citenamefont
		{Khuntia}, \citenamefont {Velazquez}, \citenamefont {Barthélemy},
		\citenamefont {Bert}, \citenamefont {Kermarrec}, \citenamefont {Legros},
		\citenamefont {Bernu}, \citenamefont {Messio}, \citenamefont {Zorko},\ and\
		\citenamefont {Mendels}}]{Khuntia469}%
	\BibitemOpen
	\bibfield  {author} {\bibinfo {author} {\bibfnamefont {P.}~\bibnamefont
			{Khuntia}}, \bibinfo {author} {\bibfnamefont {M.}~\bibnamefont {Velazquez}},
		\bibinfo {author} {\bibfnamefont {Q.}~\bibnamefont {Barthélemy}}, \bibinfo
		{author} {\bibfnamefont {F.}~\bibnamefont {Bert}}, \bibinfo {author}
		{\bibfnamefont {E.}~\bibnamefont {Kermarrec}}, \bibinfo {author}
		{\bibfnamefont {A.}~\bibnamefont {Legros}}, \bibinfo {author} {\bibfnamefont
			{B.}~\bibnamefont {Bernu}}, \bibinfo {author} {\bibfnamefont
			{L.}~\bibnamefont {Messio}}, \bibinfo {author} {\bibfnamefont
			{A.}~\bibnamefont {Zorko}},\ and\ \bibinfo {author} {\bibfnamefont
			{P.}~\bibnamefont {Mendels}},\ }\bibfield  {title} {\bibinfo {title}
		{{Gapless ground state in the archetypal quantum kagome antiferromagnet
				ZnCu$_3$(OH)$_6$Cl$_2$}},\ }\href {https://doi.org/10.1038/s41567-020-0792-1}
	{\bibfield  {journal} {\bibinfo  {journal} {Nat. Phys.}\ }\textbf {\bibinfo
			{volume} {16}},\ \bibinfo {pages} {469} (\bibinfo {year} {2020})}\BibitemShut
	{NoStop}%
	\bibitem [{\citenamefont {Fu}\ \emph {et~al.}(2015)\citenamefont {Fu},
		\citenamefont {Ima}, \citenamefont {Han},\ and\ \citenamefont {Lee}}]{Fu655}%
	\BibitemOpen
	\bibfield  {author} {\bibinfo {author} {\bibfnamefont {M.}~\bibnamefont
			{Fu}}, \bibinfo {author} {\bibfnamefont {T.}~\bibnamefont {Ima}}, \bibinfo
		{author} {\bibfnamefont {T.~H.}\ \bibnamefont {Han}},\ and\ \bibinfo {author}
		{\bibfnamefont {Y.~S.}\ \bibnamefont {Lee}},\ }\bibfield  {title} {\bibinfo
		{title} {{Evidence for a gapped spin-liquid ground state in a kagome
				Heisenberg antiferromagnet}},\ }\href
	{https://doi.org/10.1126/science.aab2120} {\bibfield  {journal} {\bibinfo
			{journal} {Science}\ }\textbf {\bibinfo {volume} {350}},\ \bibinfo {pages}
		{655} (\bibinfo {year} {2015})}\BibitemShut {NoStop}%
	\bibitem [{\citenamefont {Bieri}\ \emph {et~al.}(2015)\citenamefont {Bieri},
		\citenamefont {Messio}, \citenamefont {Bernu},\ and\ \citenamefont
		{Lhuillier}}]{Bieri060407}%
	\BibitemOpen
	\bibfield  {author} {\bibinfo {author} {\bibfnamefont {S.}~\bibnamefont
			{Bieri}}, \bibinfo {author} {\bibfnamefont {L.}~\bibnamefont {Messio}},
		\bibinfo {author} {\bibfnamefont {B.}~\bibnamefont {Bernu}},\ and\ \bibinfo
		{author} {\bibfnamefont {C.}~\bibnamefont {Lhuillier}},\ }\bibfield  {title}
	{\bibinfo {title} {{Gapless chiral spin liquid in a kagome Heisenberg
				model}},\ }\href {https://doi.org/10.1103/PhysRevB.92.060407} {\bibfield
		{journal} {\bibinfo  {journal} {Phys. Rev. B}\ }\textbf {\bibinfo {volume}
			{92}},\ \bibinfo {pages} {060407(R)} (\bibinfo {year} {2015})}\BibitemShut
	{NoStop}%
	\bibitem [{\citenamefont {Picot}\ \emph {et~al.}(2016)\citenamefont {Picot},
		\citenamefont {Ziegler}, \citenamefont {Or\'us},\ and\ \citenamefont
		{Poilblanc}}]{Picot060407}%
	\BibitemOpen
	\bibfield  {author} {\bibinfo {author} {\bibfnamefont {T.}~\bibnamefont
			{Picot}}, \bibinfo {author} {\bibfnamefont {M.}~\bibnamefont {Ziegler}},
		\bibinfo {author} {\bibfnamefont {R.}~\bibnamefont {Or\'us}},\ and\ \bibinfo
		{author} {\bibfnamefont {D.}~\bibnamefont {Poilblanc}},\ }\bibfield  {title}
	{\bibinfo {title} {{Spin-$S$ kagome quantum antiferromagnets in a field with
				tensor networks}},\ }\href {https://doi.org/10.1103/PhysRevB.93.060407}
	{\bibfield  {journal} {\bibinfo  {journal} {Phys. Rev. B}\ }\textbf {\bibinfo
			{volume} {93}},\ \bibinfo {pages} {060407} (\bibinfo {year}
		{2016})}\BibitemShut {NoStop}%
	\bibitem [{\citenamefont {Suttner}\ \emph {et~al.}(2014)\citenamefont
		{Suttner}, \citenamefont {Platt}, \citenamefont {Reuther},\ and\
		\citenamefont {Thomale}}]{Suttner020408}%
	\BibitemOpen
	\bibfield  {author} {\bibinfo {author} {\bibfnamefont {R.}~\bibnamefont
			{Suttner}}, \bibinfo {author} {\bibfnamefont {C.}~\bibnamefont {Platt}},
		\bibinfo {author} {\bibfnamefont {J.}~\bibnamefont {Reuther}},\ and\ \bibinfo
		{author} {\bibfnamefont {R.}~\bibnamefont {Thomale}},\ }\bibfield  {title}
	{\bibinfo {title} {{Renormalization group analysis of competing quantum
				phases in the ${J}_{1}$-${J}_{2}$ Heisenberg model on the kagome lattice}},\
	}\href {https://doi.org/10.1103/PhysRevB.89.020408} {\bibfield  {journal}
		{\bibinfo  {journal} {Phys. Rev. B}\ }\textbf {\bibinfo {volume} {89}},\
		\bibinfo {pages} {020408} (\bibinfo {year} {2014})}\BibitemShut {NoStop}%
	\bibitem [{\citenamefont {Hering}\ \emph {et~al.}(2022)\citenamefont {Hering},
		\citenamefont {Ferrari}, \citenamefont {Razpopov}, \citenamefont {Mazin},
		\citenamefont {Valenti}, \citenamefont {Jeschke},\ and\ \citenamefont
		{Reuther}}]{Hering10}%
	\BibitemOpen
	\bibfield  {author} {\bibinfo {author} {\bibfnamefont {M.}~\bibnamefont
			{Hering}}, \bibinfo {author} {\bibfnamefont {F.}~\bibnamefont {Ferrari}},
		\bibinfo {author} {\bibfnamefont {A.}~\bibnamefont {Razpopov}}, \bibinfo
		{author} {\bibfnamefont {I.~I.}\ \bibnamefont {Mazin}}, \bibinfo {author}
		{\bibfnamefont {R.}~\bibnamefont {Valenti}}, \bibinfo {author} {\bibfnamefont
			{H.~O.}\ \bibnamefont {Jeschke}},\ and\ \bibinfo {author} {\bibfnamefont
			{J.}~\bibnamefont {Reuther}},\ }\bibfield  {title} {\bibinfo {title} {{Phase
				diagram of a distorted kagome antiferromagnet and application to
				Y-kapellasite}},\ }\href {https://doi.org/10.1038/s41524-021-00689-0}
	{\bibfield  {journal} {\bibinfo  {journal} {Npj Comput. Mater}\ }\textbf
		{\bibinfo {volume} {8}},\ \bibinfo {pages} {10} (\bibinfo {year}
		{2022})}\BibitemShut {NoStop}%
	\bibitem [{\citenamefont {Fujihala}\ \emph {et~al.}(2020)\citenamefont
		{Fujihala}, \citenamefont {Morita}, \citenamefont {Mole}, \citenamefont
		{Mitsuda}, \citenamefont {Tohyama}, \citenamefont {Yano} \emph
		{et~al.}}]{Fujihala3429}%
	\BibitemOpen
	\bibfield  {author} {\bibinfo {author} {\bibfnamefont {M.}~\bibnamefont
			{Fujihala}}, \bibinfo {author} {\bibfnamefont {K.}~\bibnamefont {Morita}},
		\bibinfo {author} {\bibfnamefont {R.}~\bibnamefont {Mole}}, \bibinfo {author}
		{\bibfnamefont {S.}~\bibnamefont {Mitsuda}}, \bibinfo {author} {\bibfnamefont
			{T.}~\bibnamefont {Tohyama}}, \bibinfo {author} {\bibfnamefont
			{S.}~\bibnamefont {Yano}}, \emph {et~al.},\ }\bibfield  {title} {\bibinfo
		{title} {{Gapless spin liquid in a square-kagome lattice antiferromagnet}},\
	}\href {https://doi.org/10.1038/s41467-020-17235-z} {\bibfield  {journal}
		{\bibinfo  {journal} {Nat. Commun.}\ }\textbf {\bibinfo {volume} {11}},\
		\bibinfo {pages} {3429} (\bibinfo {year} {2020})}\BibitemShut {NoStop}%
	\bibitem [{\citenamefont {Tang}\ \emph {et~al.}(2017)\citenamefont {Tang},
		\citenamefont {Peng}, \citenamefont {Guo}, \citenamefont {Wang},
		\citenamefont {Su},\ and\ \citenamefont {He}}]{Tang14057}%
	\BibitemOpen
	\bibfield  {author} {\bibinfo {author} {\bibfnamefont {Y.~Y.}\ \bibnamefont
			{Tang}}, \bibinfo {author} {\bibfnamefont {C.}~\bibnamefont {Peng}}, \bibinfo
		{author} {\bibfnamefont {W.~B.}\ \bibnamefont {Guo}}, \bibinfo {author}
		{\bibfnamefont {J.~F.}\ \bibnamefont {Wang}}, \bibinfo {author}
		{\bibfnamefont {G.}~\bibnamefont {Su}},\ and\ \bibinfo {author}
		{\bibfnamefont {Z.~Z.}\ \bibnamefont {He}},\ }\bibfield  {title} {\bibinfo
		{title} {{Octa-kagome lattice compounds showing quantum critical behaviours:
				Spin gap ground state versus antiferromagnetic ordering}},\ }\href
	{https://doi.org/10.1021/jacs.7b09246} {\bibfield  {journal} {\bibinfo
			{journal} {J. Am. Chem. Soc.}\ }\textbf {\bibinfo {volume} {139}},\ \bibinfo
		{pages} {14057} (\bibinfo {year} {2017})}\BibitemShut {NoStop}%
	\bibitem [{\citenamefont {Morosan}\ \emph {et~al.}(2007)\citenamefont
		{Morosan}, \citenamefont {Fleitman}, \citenamefont {Klimczuk},\ and\
		\citenamefont {Cava}}]{Morosan144403}%
	\BibitemOpen
	\bibfield  {author} {\bibinfo {author} {\bibfnamefont {E.}~\bibnamefont
			{Morosan}}, \bibinfo {author} {\bibfnamefont {J.}~\bibnamefont {Fleitman}},
		\bibinfo {author} {\bibfnamefont {T.}~\bibnamefont {Klimczuk}},\ and\
		\bibinfo {author} {\bibfnamefont {R.~J.}\ \bibnamefont {Cava}},\ }\bibfield
	{title} {\bibinfo {title} {{Rich magnetic phase diagram of the
				kagome-staircase compound Mn$_3$V$_2$O$_8$}},\ }\href
	{https://doi.org/10.1103/PhysRevB.76.144403} {\bibfield  {journal} {\bibinfo
			{journal} {Phys. Rev. B}\ }\textbf {\bibinfo {volume} {76}},\ \bibinfo
		{pages} {144403} (\bibinfo {year} {2007})}\BibitemShut {NoStop}%
	\bibitem [{\citenamefont {Yoo}\ \emph {et~al.}(2018)\citenamefont {Yoo},
		\citenamefont {Koteswararao}, \citenamefont {Kang}, \citenamefont {Shahee}
		\emph {et~al.}}]{Yoo2397}%
	\BibitemOpen
	\bibfield  {author} {\bibinfo {author} {\bibfnamefont {K.}~\bibnamefont
			{Yoo}}, \bibinfo {author} {\bibfnamefont {B.}~\bibnamefont {Koteswararao}},
		\bibinfo {author} {\bibfnamefont {J.}~\bibnamefont {Kang}}, \bibinfo {author}
		{\bibfnamefont {A.}~\bibnamefont {Shahee}}, \emph {et~al.},\ }\bibfield
	{title} {\bibinfo {title} {{Magnetic field-induced ferroelectricity in
				spin-$\frac{1}{2}$ kagome staircase compound PbCu$_3$TeO$_7$}},\ }\href
	{https://doi.org/10.1038/s41535-018-0117-0} {\bibfield  {journal} {\bibinfo
			{journal} {npj Quantum Mater.}\ }\textbf {\bibinfo {volume} {3}},\ \bibinfo
		{pages} {2397} (\bibinfo {year} {2018})}\BibitemShut {NoStop}%
	\bibitem [{\citenamefont {Rousochatzakis}\ \emph {et~al.}(2008)\citenamefont
		{Rousochatzakis}, \citenamefont {Lauchli},\ and\ \citenamefont
		{Mila}}]{Rousochatzakis094420}%
	\BibitemOpen
	\bibfield  {author} {\bibinfo {author} {\bibfnamefont {I.}~\bibnamefont
			{Rousochatzakis}}, \bibinfo {author} {\bibfnamefont {A.~M.}\ \bibnamefont
			{Lauchli}},\ and\ \bibinfo {author} {\bibfnamefont {F.}~\bibnamefont
			{Mila}},\ }\bibfield  {title} {\bibinfo {title} {Highly frustrated magnetic
			clusters: The kagome on a sphere},\ }\href
	{https://doi.org/10.1103/PhysRevB.77.094420} {\bibfield  {journal} {\bibinfo
			{journal} {Phys. Rev. B}\ }\textbf {\bibinfo {volume} {77}},\ \bibinfo
		{pages} {094420} (\bibinfo {year} {2008})}\BibitemShut {NoStop}%
	\bibitem [{\citenamefont {Jeschke}\ \emph {et~al.}(2019)\citenamefont
		{Jeschke}, \citenamefont {Nakano},\ and\ \citenamefont
		{Sakai}}]{Jeschke140410}%
	\BibitemOpen
	\bibfield  {author} {\bibinfo {author} {\bibfnamefont {H.~O.}\ \bibnamefont
			{Jeschke}}, \bibinfo {author} {\bibfnamefont {H.}~\bibnamefont {Nakano}},\
		and\ \bibinfo {author} {\bibfnamefont {T.}~\bibnamefont {Sakai}},\ }\bibfield
	{title} {\bibinfo {title} {{From kagome strip to kagome lattice:
				Realizations of frustrated spin-$\frac{1}{2}$ antiferromagnets in Ti(III)
				fluorides}},\ }\href {https://doi.org/10.1103/PhysRevB.99.140410} {\bibfield
		{journal} {\bibinfo  {journal} {Phys. Rev. B}\ }\textbf {\bibinfo {volume}
			{99}},\ \bibinfo {pages} {140410} (\bibinfo {year} {2019})}\BibitemShut
	{NoStop}%
	\bibitem [{\citenamefont {Morita}\ \emph {et~al.}(2021)\citenamefont {Morita},
		\citenamefont {Sota},\ and\ \citenamefont {Tohyama}}]{Morita2399}%
	\BibitemOpen
	\bibfield  {author} {\bibinfo {author} {\bibfnamefont {K.}~\bibnamefont
			{Morita}}, \bibinfo {author} {\bibfnamefont {S.}~\bibnamefont {Sota}},\ and\
		\bibinfo {author} {\bibfnamefont {T.}~\bibnamefont {Tohyama}},\ }\bibfield
	{title} {\bibinfo {title} {{Resonating dimer-monomer liquid state in a
				magnetization plateau of a spin-$\frac{1}{2}$ kagome-strip Heisenberg
				chain}},\ }\href {https://doi.org/10.1038/s42005-021-00665-6} {\bibfield
		{journal} {\bibinfo  {journal} {Commun. Phys.}\ }\textbf {\bibinfo {volume}
			{4}},\ \bibinfo {pages} {2399} (\bibinfo {year} {2021})}\BibitemShut
	{NoStop}%
	\bibitem [{\citenamefont {Dun}\ \emph {et~al.}(2016)\citenamefont {Dun},
		\citenamefont {Trinh}, \citenamefont {Li}, \citenamefont {Lee}, \citenamefont
		{Chen}, \citenamefont {Baumbach}, \citenamefont {Hu} \emph
		{et~al.}}]{Dun157201}%
	\BibitemOpen
	\bibfield  {author} {\bibinfo {author} {\bibfnamefont {Z.~L.}\ \bibnamefont
			{Dun}}, \bibinfo {author} {\bibfnamefont {J.}~\bibnamefont {Trinh}}, \bibinfo
		{author} {\bibfnamefont {K.}~\bibnamefont {Li}}, \bibinfo {author}
		{\bibfnamefont {M.}~\bibnamefont {Lee}}, \bibinfo {author} {\bibfnamefont
			{K.~W.}\ \bibnamefont {Chen}}, \bibinfo {author} {\bibfnamefont
			{R.}~\bibnamefont {Baumbach}}, \bibinfo {author} {\bibfnamefont {Y.~F.}\
			\bibnamefont {Hu}}, \emph {et~al.},\ }\bibfield  {title} {\bibinfo {title}
		{{Magnetic Ground States of the Rare-Earth Tripod Kagome Lattice
				${\mathrm{Mg}}_{2}{\mathrm{RE}}_{3}{\mathrm{Sb}}_{3}{\mathrm{O}}_{14}$
				($\mathrm{RE}=\mathrm{Gd},\mathrm{Dy},\mathrm{Er}$)}},\ }\href
	{https://doi.org/10.1103/PhysRevLett.116.157201} {\bibfield  {journal}
		{\bibinfo  {journal} {Phys. Rev. Lett.}\ }\textbf {\bibinfo {volume} {116}},\
		\bibinfo {pages} {157201} (\bibinfo {year} {2016})}\BibitemShut {NoStop}%
	\bibitem [{\citenamefont {Dun}\ \emph {et~al.}(2020)\citenamefont {Dun},
		\citenamefont {Bai}, \citenamefont {Paddison}, \citenamefont {Hollingworth},
		\citenamefont {Butch}, \citenamefont {Cruz}, \citenamefont {Stone} \emph
		{et~al.}}]{Dun031069}%
	\BibitemOpen
	\bibfield  {author} {\bibinfo {author} {\bibfnamefont {Z.}~\bibnamefont
			{Dun}}, \bibinfo {author} {\bibfnamefont {X.}~\bibnamefont {Bai}}, \bibinfo
		{author} {\bibfnamefont {J.~A.~M.}\ \bibnamefont {Paddison}}, \bibinfo
		{author} {\bibfnamefont {E.}~\bibnamefont {Hollingworth}}, \bibinfo {author}
		{\bibfnamefont {N.~P.}\ \bibnamefont {Butch}}, \bibinfo {author}
		{\bibfnamefont {C.~D.}\ \bibnamefont {Cruz}}, \bibinfo {author}
		{\bibfnamefont {M.~B.}\ \bibnamefont {Stone}}, \emph {et~al.},\ }\bibfield
	{title} {\bibinfo {title} {{Quantum Versus Classical Spin Fragmentation in
				Dipolar Kagome Ice
				${\mathrm{Ho}}_{3}{\mathrm{Mg}}_{2}{\mathrm{Sb}}_{3}{\mathrm{O}}_{14}$}},\
	}\href {https://doi.org/10.1103/PhysRevX.10.031069} {\bibfield  {journal}
		{\bibinfo  {journal} {Phys. Rev. X}\ }\textbf {\bibinfo {volume} {10}},\
		\bibinfo {pages} {031069} (\bibinfo {year} {2020})}\BibitemShut {NoStop}%
	\bibitem [{\citenamefont {Okamoto}\ \emph {et~al.}(2007)\citenamefont
		{Okamoto}, \citenamefont {Nohara}, \citenamefont {Aruga-Katori},\ and\
		\citenamefont {Takagi}}]{Okamoto137207}%
	\BibitemOpen
	\bibfield  {author} {\bibinfo {author} {\bibfnamefont {Y.}~\bibnamefont
			{Okamoto}}, \bibinfo {author} {\bibfnamefont {M.}~\bibnamefont {Nohara}},
		\bibinfo {author} {\bibfnamefont {H.}~\bibnamefont {Aruga-Katori}},\ and\
		\bibinfo {author} {\bibfnamefont {H.}~\bibnamefont {Takagi}},\ }\bibfield
	{title} {\bibinfo {title} {{Spin-Liquid State in the spin-$\frac{1}{2}$
				Hyperkagome Antiferromagnet
				${\mathrm{Na}}_{4}{\mathrm{Ir}}_{3}{\mathrm{O}}_{8}$}},\ }\href
	{https://doi.org/10.1103/PhysRevLett.99.137207} {\bibfield  {journal}
		{\bibinfo  {journal} {Phys. Rev. Lett.}\ }\textbf {\bibinfo {volume} {99}},\
		\bibinfo {pages} {137207} (\bibinfo {year} {2007})}\BibitemShut {NoStop}%
	\bibitem [{\citenamefont {Lawler}\ \emph {et~al.}(2008)\citenamefont {Lawler},
		\citenamefont {Kee}, \citenamefont {Kim},\ and\ \citenamefont
		{Vishwanath}}]{Lawler227201}%
	\BibitemOpen
	\bibfield  {author} {\bibinfo {author} {\bibfnamefont {M.~J.}\ \bibnamefont
			{Lawler}}, \bibinfo {author} {\bibfnamefont {H.~Y.}\ \bibnamefont {Kee}},
		\bibinfo {author} {\bibfnamefont {Y.~B.}\ \bibnamefont {Kim}},\ and\ \bibinfo
		{author} {\bibfnamefont {A.}~\bibnamefont {Vishwanath}},\ }\bibfield  {title}
	{\bibinfo {title} {{Topological Spin Liquid on the Hyperkagome Lattice of
				${\mathrm{Na}}_{4}{\mathrm{Ir}}_{3}{\mathrm{O}}_{8}$}},\ }\href
	{https://doi.org/10.1103/PhysRevLett.100.227201} {\bibfield  {journal}
		{\bibinfo  {journal} {Phys. Rev. Lett.}\ }\textbf {\bibinfo {volume} {100}},\
		\bibinfo {pages} {227201} (\bibinfo {year} {2008})}\BibitemShut {NoStop}%
	\bibitem [{\citenamefont {Botana}\ \emph {et~al.}(2018)\citenamefont {Botana},
		\citenamefont {Zheng}, \citenamefont {Lapidus}, \citenamefont {Mitchell},\
		and\ \citenamefont {Norman}}]{Botana054421}%
	\BibitemOpen
	\bibfield  {author} {\bibinfo {author} {\bibfnamefont {A.~S.}\ \bibnamefont
			{Botana}}, \bibinfo {author} {\bibfnamefont {H.}~\bibnamefont {Zheng}},
		\bibinfo {author} {\bibfnamefont {S.~H.}\ \bibnamefont {Lapidus}}, \bibinfo
		{author} {\bibfnamefont {J.~F.}\ \bibnamefont {Mitchell}},\ and\ \bibinfo
		{author} {\bibfnamefont {M.~R.}\ \bibnamefont {Norman}},\ }\bibfield  {title}
	{\bibinfo {title} {Averievite: A copper oxide kagome antiferromagnet},\
	}\href {https://doi.org/10.1103/PhysRevB.98.054421} {\bibfield  {journal}
		{\bibinfo  {journal} {Phys. Rev. B}\ }\textbf {\bibinfo {volume} {98}},\
		\bibinfo {pages} {054421} (\bibinfo {year} {2018})}\BibitemShut {NoStop}%
	\bibitem [{\citenamefont {Dey}\ and\ \citenamefont {Botana}(2020)}]{Dey125106}%
	\BibitemOpen
	\bibfield  {author} {\bibinfo {author} {\bibfnamefont {D.}~\bibnamefont
			{Dey}}\ and\ \bibinfo {author} {\bibfnamefont {A.~S.}\ \bibnamefont
			{Botana}},\ }\bibfield  {title} {\bibinfo {title} {Role of chemical pressure
			on the electronic and magnetic properties of the spin-$\frac{1}{2}$ kagome
			mineral averievite},\ }\href {https://doi.org/10.1103/PhysRevB.102.125106}
	{\bibfield  {journal} {\bibinfo  {journal} {Phys. Rev. B}\ }\textbf {\bibinfo
			{volume} {102}},\ \bibinfo {pages} {125106} (\bibinfo {year}
		{2020})}\BibitemShut {NoStop}%
	\bibitem [{\citenamefont {Georgopoulou}\ \emph {et~al.}()\citenamefont
		{Georgopoulou}, \citenamefont {Boldrin}, \citenamefont {Fak}, \citenamefont
		{Manuel}, \citenamefont {Gibbs}, \citenamefont {Ollivier}, \citenamefont
		{Suard},\ and\ \citenamefont {Wills}}]{Georgopoulou14742}%
	\BibitemOpen
	\bibfield  {author} {\bibinfo {author} {\bibfnamefont {M.}~\bibnamefont
			{Georgopoulou}}, \bibinfo {author} {\bibfnamefont {D.}~\bibnamefont
			{Boldrin}}, \bibinfo {author} {\bibfnamefont {B.}~\bibnamefont {Fak}},
		\bibinfo {author} {\bibfnamefont {P.}~\bibnamefont {Manuel}}, \bibinfo
		{author} {\bibfnamefont {A.}~\bibnamefont {Gibbs}}, \bibinfo {author}
		{\bibfnamefont {J.}~\bibnamefont {Ollivier}}, \bibinfo {author}
		{\bibfnamefont {E.}~\bibnamefont {Suard}},\ and\ \bibinfo {author}
		{\bibfnamefont {A.~S.}\ \bibnamefont {Wills}},\ }\bibfield  {title} {\bibinfo
		{title} {Magnetic ground states and excitations in {Zn}-doped averieite - a
			family of oxide-based spin-$\frac{1}{2}$ kagome antiferromagnets},\ }\href
	{https://doi.org/https://doi.org/10.48550/arXiv.2306.14739} {\bibinfo
		{journal} {arXiv:2306.14739v1}\ }\BibitemShut {NoStop}%
	\bibitem [{\citenamefont {Biesner}\ \emph {et~al.}(2022)\citenamefont
		{Biesner}, \citenamefont {Roh}, \citenamefont {Pustogow}, \citenamefont
		{Zheng}, \citenamefont {Mitchell},\ and\ \citenamefont
		{Dressel}}]{BiesnerL060410}%
	\BibitemOpen
	\bibfield  {journal} {  }\bibfield  {author} {\bibinfo {author} {\bibfnamefont
			{T.}~\bibnamefont {Biesner}}, \bibinfo {author} {\bibfnamefont
			{S.}~\bibnamefont {Roh}}, \bibinfo {author} {\bibfnamefont {A.}~\bibnamefont
			{Pustogow}}, \bibinfo {author} {\bibfnamefont {H.}~\bibnamefont {Zheng}},
		\bibinfo {author} {\bibfnamefont {J.~F.}\ \bibnamefont {Mitchell}},\ and\
		\bibinfo {author} {\bibfnamefont {M.}~\bibnamefont {Dressel}},\ }\bibfield
	{title} {\bibinfo {title} {Magnetic terahertz resonances above the n\'eel
			temperature in the frustrated kagome antiferromagnet averievite},\ }\href
	{https://doi.org/10.1103/PhysRevB.105.L060410} {\bibfield  {journal}
		{\bibinfo  {journal} {Phys. Rev. B}\ }\textbf {\bibinfo {volume} {105}},\
		\bibinfo {pages} {L060410} (\bibinfo {year} {2022})}\BibitemShut {NoStop}%
	\bibitem [{\citenamefont {Winiarski}\ \emph {et~al.}(2019)\citenamefont
		{Winiarski}, \citenamefont {Tran}, \citenamefont {Chamorro},\ and\
		\citenamefont {McQueen}}]{Winiarski4328}%
	\BibitemOpen
	\bibfield  {author} {\bibinfo {author} {\bibfnamefont {M.~J.}\ \bibnamefont
			{Winiarski}}, \bibinfo {author} {\bibfnamefont {T.~T.}\ \bibnamefont {Tran}},
		\bibinfo {author} {\bibfnamefont {J.~R.}\ \bibnamefont {Chamorro}},\ and\
		\bibinfo {author} {\bibfnamefont {T.~M.}\ \bibnamefont {McQueen}},\
	}\bibfield  {title} {\bibinfo {title} {{(Cs$X$)Cu$_5$O$_2$(PO$_4$)$_2$ ($X$ =
				Cl, Br, I): A Family of Cu$^{2+}$ spin-$\frac{1}{2}$ Compounds with
				Capped-Kagomé Networks Composed of OCu$_4$ Units}},\ }\href
	{https://doi.org/10.1021/acs.inorgchem.8b03464} {\bibfield  {journal}
		{\bibinfo  {journal} {Inorg. Chem.}\ }\textbf {\bibinfo {volume} {58}},\
		\bibinfo {pages} {4328} (\bibinfo {year} {2019})}\BibitemShut {NoStop}%
	\bibitem [{\citenamefont {Zhang}\ \emph {et~al.}(2020)\citenamefont {Zhang},
		\citenamefont {He}, \citenamefont {Xie}, \citenamefont {Cui}, \citenamefont
		{Zhang}, \citenamefont {Chen}, \citenamefont {Zhao}, \citenamefont {Zhang},\
		and\ \citenamefont {Huang}}]{Zhang2299}%
	\BibitemOpen
	\bibfield  {author} {\bibinfo {author} {\bibfnamefont {W.}~\bibnamefont
			{Zhang}}, \bibinfo {author} {\bibfnamefont {Z.}~\bibnamefont {He}}, \bibinfo
		{author} {\bibfnamefont {Y.}~\bibnamefont {Xie}}, \bibinfo {author}
		{\bibfnamefont {M.}~\bibnamefont {Cui}}, \bibinfo {author} {\bibfnamefont
			{S.}~\bibnamefont {Zhang}}, \bibinfo {author} {\bibfnamefont
			{S.}~\bibnamefont {Chen}}, \bibinfo {author} {\bibfnamefont {Z.}~\bibnamefont
			{Zhao}}, \bibinfo {author} {\bibfnamefont {M.}~\bibnamefont {Zhang}},\ and\
		\bibinfo {author} {\bibfnamefont {X.}~\bibnamefont {Huang}},\ }\bibfield
	{title} {\bibinfo {title} {Molybdate–tellurite compounds with capped-kagome
			spin–lattices},\ }\href {https://doi.org/10.1021/acs.inorgchem.9b03050}
	{\bibfield  {journal} {\bibinfo  {journal} {Inorg. Chem.}\ }\textbf {\bibinfo
			{volume} {59}},\ \bibinfo {pages} {2299} (\bibinfo {year}
		{2020})}\BibitemShut {NoStop}%
	\bibitem [{\citenamefont {Kornyakov}\ \emph {et~al.}(2021)\citenamefont
		{Kornyakov}, \citenamefont {Vladimirova}, \citenamefont {Siidra},\ and\
		\citenamefont {Krivovichev}}]{Kornyakov1833}%
	\BibitemOpen
	\bibfield  {author} {\bibinfo {author} {\bibfnamefont {I.~V.}\ \bibnamefont
			{Kornyakov}}, \bibinfo {author} {\bibfnamefont {V.~A.}\ \bibnamefont
			{Vladimirova}}, \bibinfo {author} {\bibfnamefont {O.~I.}\ \bibnamefont
			{Siidra}},\ and\ \bibinfo {author} {\bibfnamefont {S.~V.}\ \bibnamefont
			{Krivovichev}},\ }\bibfield  {title} {\bibinfo {title} {{Expanding the
				Averievite Family, ($MX$)Cu$_5$O$_2$($T$$^{5+}$O$_4$)$_2$ ($T$$^{5+}$=P,V;
				$M$=K,Rb,Cs,Cu; $X$=Cl,Br): Synthesis and Single-Crystal X-ray Diffraction
				Study}},\ }\href {https://doi.org/10.3390/molecules26071833} {\bibfield
		{journal} {\bibinfo  {journal} {Molecules}\ }\textbf {\bibinfo {volume}
			{26}},\ \bibinfo {pages} {1833} (\bibinfo {year} {2021})}\BibitemShut
	{NoStop}%
	\bibitem [{\citenamefont {Carvajal}(1993)}]{Carvajal55}%
	\BibitemOpen
	\bibfield  {author} {\bibinfo {author} {\bibfnamefont {J.~R.}\ \bibnamefont
			{Carvajal}},\ }\bibfield  {title} {\bibinfo {title} {Recent advances in
			magnetic structure determination by neutron powder diffraction},\ }\href
	{https://doi.org/https://doi.org/10.1016/0921-4526(93)90108-I} {\bibfield
		{journal} {\bibinfo  {journal} {Physica B: Condens. Matter}\ }\textbf
		{\bibinfo {volume} {192}},\ \bibinfo {pages} {55} (\bibinfo {year}
		{1993})}\BibitemShut {NoStop}%
	\bibitem [{\citenamefont {Bader}\ \emph {et~al.}(2022)\citenamefont {Bader},
		\citenamefont {Langmann}, \citenamefont {Gegenwart},\ and\ \citenamefont
		{Tsirlin}}]{Bader054415}%
	\BibitemOpen
	\bibfield  {author} {\bibinfo {author} {\bibfnamefont {V.~P.}\ \bibnamefont
			{Bader}}, \bibinfo {author} {\bibfnamefont {J.}~\bibnamefont {Langmann}},
		\bibinfo {author} {\bibfnamefont {P.}~\bibnamefont {Gegenwart}},\ and\
		\bibinfo {author} {\bibfnamefont {A.~A.}\ \bibnamefont {Tsirlin}},\
	}\bibfield  {title} {\bibinfo {title} {{Deformation of the triangular
				spin-$\frac{1}{2}$ lattice in
				${\mathrm{Na}}_{2}{\mathrm{SrCo}(\mathrm{PO}}_{4}{)}_{2}$}},\ }\href
	{https://doi.org/10.1103/PhysRevB.106.054415} {\bibfield  {journal} {\bibinfo
			{journal} {Phys. Rev. B}\ }\textbf {\bibinfo {volume} {106}},\ \bibinfo
		{pages} {054415} (\bibinfo {year} {2022})}\BibitemShut {NoStop}%
	\bibitem [{\citenamefont {Tsirlin}\ \emph {et~al.}(2011)\citenamefont
		{Tsirlin}, \citenamefont {Nath}, \citenamefont {Abakumov}, \citenamefont
		{Furukawa}, \citenamefont {Johnston}, \citenamefont {Hemmida}, \citenamefont
		{Krug~von Nidda}, \citenamefont {Loidl}, \citenamefont {Geibel},\ and\
		\citenamefont {Rosner}}]{Tsirlin014429}%
	\BibitemOpen
	\bibfield  {author} {\bibinfo {author} {\bibfnamefont {A.~A.}\ \bibnamefont
			{Tsirlin}}, \bibinfo {author} {\bibfnamefont {R.}~\bibnamefont {Nath}},
		\bibinfo {author} {\bibfnamefont {A.~M.}\ \bibnamefont {Abakumov}}, \bibinfo
		{author} {\bibfnamefont {Y.}~\bibnamefont {Furukawa}}, \bibinfo {author}
		{\bibfnamefont {D.~C.}\ \bibnamefont {Johnston}}, \bibinfo {author}
		{\bibfnamefont {M.}~\bibnamefont {Hemmida}}, \bibinfo {author} {\bibfnamefont
			{H.-A.}\ \bibnamefont {Krug~von Nidda}}, \bibinfo {author} {\bibfnamefont
			{A.}~\bibnamefont {Loidl}}, \bibinfo {author} {\bibfnamefont
			{C.}~\bibnamefont {Geibel}},\ and\ \bibinfo {author} {\bibfnamefont
			{H.}~\bibnamefont {Rosner}},\ }\bibfield  {title} {\bibinfo {title} {{Phase
				separation and frustrated square lattice magnetism of
				Na${}_{1.5}$VOPO${}_{4}$F${}_{0.5}$}},\ }\href
	{https://doi.org/10.1103/PhysRevB.84.014429} {\bibfield  {journal} {\bibinfo
			{journal} {Phys. Rev. B}\ }\textbf {\bibinfo {volume} {84}},\ \bibinfo
		{pages} {014429} (\bibinfo {year} {2011})}\BibitemShut {NoStop}%
	\bibitem [{\citenamefont {Guchhait}\ \emph {et~al.}(2022)\citenamefont
		{Guchhait}, \citenamefont {Ambika}, \citenamefont {Ding}, \citenamefont
		{Uhlarz}, \citenamefont {Furukawa}, \citenamefont {Tsirlin},\ and\
		\citenamefont {Nath}}]{Guchhait024426}%
	\BibitemOpen
	\bibfield  {author} {\bibinfo {author} {\bibfnamefont {S.}~\bibnamefont
			{Guchhait}}, \bibinfo {author} {\bibfnamefont {D.~V.}\ \bibnamefont
			{Ambika}}, \bibinfo {author} {\bibfnamefont {Q.-P.}\ \bibnamefont {Ding}},
		\bibinfo {author} {\bibfnamefont {M.}~\bibnamefont {Uhlarz}}, \bibinfo
		{author} {\bibfnamefont {Y.}~\bibnamefont {Furukawa}}, \bibinfo {author}
		{\bibfnamefont {A.~A.}\ \bibnamefont {Tsirlin}},\ and\ \bibinfo {author}
		{\bibfnamefont {R.}~\bibnamefont {Nath}},\ }\bibfield  {title} {\bibinfo
		{title} {{Deformed spin-$\frac{1}{2}$ square lattice in antiferromagnetic
				${\mathrm{NaZnVOPO}}_{4}({\mathrm{HPO}}_{4})$}},\ }\href
	{https://doi.org/10.1103/PhysRevB.106.024426} {\bibfield  {journal} {\bibinfo
			{journal} {Phys. Rev. B}\ }\textbf {\bibinfo {volume} {106}},\ \bibinfo
		{pages} {024426} (\bibinfo {year} {2022})}\BibitemShut {NoStop}%
	\bibitem [{\citenamefont {Kittel}(2004)}]{Kittel2004}%
	\BibitemOpen
	\bibfield  {author} {\bibinfo {author} {\bibfnamefont {C.}~\bibnamefont
			{Kittel}},\ }\href {https://books.google.co.in/books?id=kym4QgAACAAJ} {\emph
		{\bibinfo {title} {Introduction to Solid State Physics}}}\ (\bibinfo
	{publisher} {Wiley},\ \bibinfo {year} {2004})\BibitemShut {NoStop}%
	\bibitem [{\citenamefont {Nath}\ \emph {et~al.}(2015)\citenamefont {Nath},
		\citenamefont {Padmanabhan}, \citenamefont {Baby}, \citenamefont
		{Thirumurugan}, \citenamefont {Ehlers}, \citenamefont {Hemmida},
		\citenamefont {Krug~von Nidda},\ and\ \citenamefont {Tsirlin}}]{Nath054409}%
	\BibitemOpen
	\bibfield  {author} {\bibinfo {author} {\bibfnamefont {R.}~\bibnamefont
			{Nath}}, \bibinfo {author} {\bibfnamefont {M.}~\bibnamefont {Padmanabhan}},
		\bibinfo {author} {\bibfnamefont {S.}~\bibnamefont {Baby}}, \bibinfo {author}
		{\bibfnamefont {A.}~\bibnamefont {Thirumurugan}}, \bibinfo {author}
		{\bibfnamefont {D.}~\bibnamefont {Ehlers}}, \bibinfo {author} {\bibfnamefont
			{M.}~\bibnamefont {Hemmida}}, \bibinfo {author} {\bibfnamefont {H.-A.}\
			\bibnamefont {Krug~von Nidda}},\ and\ \bibinfo {author} {\bibfnamefont
			{A.~A.}\ \bibnamefont {Tsirlin}},\ }\bibfield  {title} {\bibinfo {title}
		{{Quasi-two-dimensional $S=\frac{1}{2}$ magnetism of
				$\mathrm{Cu[}{\mathrm{C}}_{6}{\mathrm{H}}_{2}(\text{COO}{)}_{4}\mathrm{][}{\mathrm{C}}_{2}{\mathrm{H}}_{5}{\mathrm{NH}}_{3}\mathrm{]}{}_{2}$}},\
	}\href {https://doi.org/10.1103/PhysRevB.91.054409} {\bibfield  {journal}
		{\bibinfo  {journal} {Phys. Rev. B}\ }\textbf {\bibinfo {volume} {91}},\
		\bibinfo {pages} {054409} (\bibinfo {year} {2015})}\BibitemShut {NoStop}%
	\bibitem [{\citenamefont {Selwood}(1956)}]{Selwood1956}%
	\BibitemOpen
	\bibfield  {author} {\bibinfo {author} {\bibfnamefont {P.~W.}\ \bibnamefont
			{Selwood}},\ }\href@noop {} {\emph {\bibinfo {title} {Magnetochemistry}}}\
	(\bibinfo  {publisher} {Interscience},\ \bibinfo {address} {New York},\
	\bibinfo {year} {1956})\BibitemShut {NoStop}%
	\bibitem [{\citenamefont {Bain}\ and\ \citenamefont {Berry}(2008)}]{Bain532}%
	\BibitemOpen
	\bibfield  {author} {\bibinfo {author} {\bibfnamefont {G.~A.}\ \bibnamefont
			{Bain}}\ and\ \bibinfo {author} {\bibfnamefont {J.~F.}\ \bibnamefont
			{Berry}},\ }\bibfield  {title} {\bibinfo {title} {Diamagnetic corrections and
			pascal's constants},\ }\href {https://doi.org/10.1021/ed085p532} {\bibfield
		{journal} {\bibinfo  {journal} {J. Chem. Educ.}\ }\textbf {\bibinfo {volume}
			{85}},\ \bibinfo {pages} {532} (\bibinfo {year} {2008})}\BibitemShut
	{NoStop}%
	\bibitem [{\citenamefont {Guchhait}\ \emph {et~al.}(2021)\citenamefont
		{Guchhait}, \citenamefont {Ding}, \citenamefont {Sahoo}, \citenamefont
		{Giri}, \citenamefont {Maji}, \citenamefont {Furukawa},\ and\ \citenamefont
		{Nath}}]{Guchhait224415}%
	\BibitemOpen
	\bibfield  {author} {\bibinfo {author} {\bibfnamefont {S.}~\bibnamefont
			{Guchhait}}, \bibinfo {author} {\bibfnamefont {Q.~P.}\ \bibnamefont {Ding}},
		\bibinfo {author} {\bibfnamefont {M.}~\bibnamefont {Sahoo}}, \bibinfo
		{author} {\bibfnamefont {A.}~\bibnamefont {Giri}}, \bibinfo {author}
		{\bibfnamefont {S.}~\bibnamefont {Maji}}, \bibinfo {author} {\bibfnamefont
			{Y.}~\bibnamefont {Furukawa}},\ and\ \bibinfo {author} {\bibfnamefont
			{R.}~\bibnamefont {Nath}},\ }\bibfield  {title} {\bibinfo {title}
		{{Quasi-one-dimensional uniform spin-$\frac{1}{2}$ Heisenberg antiferromagnet
				KNaCuP$_2$O$_7$ probed by $^{31}\mathrm{P}$ and $^{23}\mathrm{Na}$ NMR}},\
	}\href {https://doi.org/10.1103/PhysRevB.103.224415} {\bibfield  {journal}
		{\bibinfo  {journal} {Phys. Rev. B}\ }\textbf {\bibinfo {volume} {103}},\
		\bibinfo {pages} {224415} (\bibinfo {year} {2021})}\BibitemShut {NoStop}%
	\bibitem [{\citenamefont {Islam}\ \emph {et~al.}(2018)\citenamefont {Islam},
		\citenamefont {Ranjith}, \citenamefont {Baenitz}, \citenamefont {Skourski},
		\citenamefont {Tsirlin},\ and\ \citenamefont {Nath}}]{Islam174432}%
	\BibitemOpen
	\bibfield  {author} {\bibinfo {author} {\bibfnamefont {S.~S.}\ \bibnamefont
			{Islam}}, \bibinfo {author} {\bibfnamefont {K.~M.}\ \bibnamefont {Ranjith}},
		\bibinfo {author} {\bibfnamefont {M.}~\bibnamefont {Baenitz}}, \bibinfo
		{author} {\bibfnamefont {Y.}~\bibnamefont {Skourski}}, \bibinfo {author}
		{\bibfnamefont {A.~A.}\ \bibnamefont {Tsirlin}},\ and\ \bibinfo {author}
		{\bibfnamefont {R.}~\bibnamefont {Nath}},\ }\bibfield  {title} {\bibinfo
		{title} {{Frustration of square cupola in Sr(TiO)Cu$_4$(PO$_4$)$_4$}},\
	}\href {https://doi.org/10.1103/PhysRevB.97.174432} {\bibfield  {journal}
		{\bibinfo  {journal} {Phys. Rev. B}\ }\textbf {\bibinfo {volume} {97}},\
		\bibinfo {pages} {174432} (\bibinfo {year} {2018})}\BibitemShut {NoStop}%
	\bibitem [{\citenamefont {Islam}\ \emph {et~al.}(2020)\citenamefont {Islam},
		\citenamefont {Singh}, \citenamefont {Somesh}, \citenamefont {Mukharjee},
		\citenamefont {Jain}, \citenamefont {Yusuf},\ and\ \citenamefont
		{Nath}}]{Islam134433}%
	\BibitemOpen
	\bibfield  {author} {\bibinfo {author} {\bibfnamefont {S.~S.}\ \bibnamefont
			{Islam}}, \bibinfo {author} {\bibfnamefont {V.}~\bibnamefont {Singh}},
		\bibinfo {author} {\bibfnamefont {K.}~\bibnamefont {Somesh}}, \bibinfo
		{author} {\bibfnamefont {P.~K.}\ \bibnamefont {Mukharjee}}, \bibinfo {author}
		{\bibfnamefont {A.}~\bibnamefont {Jain}}, \bibinfo {author} {\bibfnamefont
			{S.~M.}\ \bibnamefont {Yusuf}},\ and\ \bibinfo {author} {\bibfnamefont
			{R.}~\bibnamefont {Nath}},\ }\bibfield  {title} {\bibinfo {title}
		{{Unconventional superparamagnetic behavior in the modified cubic spinel
				compound ${\mathrm{LiNi}}_{0.5}{\mathrm{Mn}}_{1.5}{\mathrm{O}}_{4}$}},\
	}\href {https://doi.org/10.1103/PhysRevB.102.134433} {\bibfield  {journal}
		{\bibinfo  {journal} {Phys. Rev. B}\ }\textbf {\bibinfo {volume} {102}},\
		\bibinfo {pages} {134433} (\bibinfo {year} {2020})}\BibitemShut {NoStop}%
	\bibitem [{\citenamefont {Mukharjee}\ \emph {et~al.}(2019)\citenamefont
		{Mukharjee}, \citenamefont {Ranjith}, \citenamefont {Koo}, \citenamefont
		{Sichelschmidt}, \citenamefont {Baenitz}, \citenamefont {Skourski},
		\citenamefont {Inagaki}, \citenamefont {Furukawa}, \citenamefont {Tsirlin},\
		and\ \citenamefont {Nath}}]{Mukharjee144433}%
	\BibitemOpen
	\bibfield  {author} {\bibinfo {author} {\bibfnamefont {P.~K.}\ \bibnamefont
			{Mukharjee}}, \bibinfo {author} {\bibfnamefont {K.~M.}\ \bibnamefont
			{Ranjith}}, \bibinfo {author} {\bibfnamefont {B.}~\bibnamefont {Koo}},
		\bibinfo {author} {\bibfnamefont {J.}~\bibnamefont {Sichelschmidt}}, \bibinfo
		{author} {\bibfnamefont {M.}~\bibnamefont {Baenitz}}, \bibinfo {author}
		{\bibfnamefont {Y.}~\bibnamefont {Skourski}}, \bibinfo {author}
		{\bibfnamefont {Y.}~\bibnamefont {Inagaki}}, \bibinfo {author} {\bibfnamefont
			{Y.}~\bibnamefont {Furukawa}}, \bibinfo {author} {\bibfnamefont {A.~A.}\
			\bibnamefont {Tsirlin}},\ and\ \bibinfo {author} {\bibfnamefont
			{R.}~\bibnamefont {Nath}},\ }\bibfield  {title} {\bibinfo {title}
		{{Bose-Einstein condensation of triplons close to the quantum critical point
				in the quasi-one-dimensional spin-$\frac{1}{2}$ antiferromagnet
				${\mathrm{NaVOPO}}_{4}$}},\ }\href
	{https://doi.org/10.1103/PhysRevB.100.144433} {\bibfield  {journal} {\bibinfo
			{journal} {Phys. Rev. B}\ }\textbf {\bibinfo {volume} {100}},\ \bibinfo
		{pages} {144433} (\bibinfo {year} {2019})}\BibitemShut {NoStop}%
	\bibitem [{\citenamefont {Yogi}\ \emph {et~al.}(2015)\citenamefont {Yogi},
		\citenamefont {Ahmed}, \citenamefont {Nath}, \citenamefont {Tsirlin},
		\citenamefont {Kundu}, \citenamefont {Mahajan}, \citenamefont
		{Sichelschmidt}, \citenamefont {Roy},\ and\ \citenamefont
		{Furukawa}}]{Yogi024413}%
	\BibitemOpen
	\bibfield  {author} {\bibinfo {author} {\bibfnamefont {A.}~\bibnamefont
			{Yogi}}, \bibinfo {author} {\bibfnamefont {N.}~\bibnamefont {Ahmed}},
		\bibinfo {author} {\bibfnamefont {R.}~\bibnamefont {Nath}}, \bibinfo {author}
		{\bibfnamefont {A.~A.}\ \bibnamefont {Tsirlin}}, \bibinfo {author}
		{\bibfnamefont {S.}~\bibnamefont {Kundu}}, \bibinfo {author} {\bibfnamefont
			{A.~V.}\ \bibnamefont {Mahajan}}, \bibinfo {author} {\bibfnamefont
			{J.}~\bibnamefont {Sichelschmidt}}, \bibinfo {author} {\bibfnamefont
			{B.}~\bibnamefont {Roy}},\ and\ \bibinfo {author} {\bibfnamefont
			{Y.}~\bibnamefont {Furukawa}},\ }\bibfield  {title} {\bibinfo {title}
		{{Antiferromagnetism of
				${\mathrm{Zn}}_{2}\mathrm{VO}{{(\mathrm{PO}}_{4})}_{2}$ and the dilution with
				${\mathrm{Ti}}^{4+}$}},\ }\href {https://doi.org/10.1103/PhysRevB.91.024413}
	{\bibfield  {journal} {\bibinfo  {journal} {Phys. Rev. B}\ }\textbf {\bibinfo
			{volume} {91}},\ \bibinfo {pages} {024413} (\bibinfo {year}
		{2015})}\BibitemShut {NoStop}%
	\bibitem [{\citenamefont {Kontani}\ \emph {et~al.}(1975)\citenamefont
		{Kontani}, \citenamefont {Hioki},\ and\ \citenamefont {Masuda}}]{Kontani672}%
	\BibitemOpen
	\bibfield  {author} {\bibinfo {author} {\bibfnamefont {M.}~\bibnamefont
			{Kontani}}, \bibinfo {author} {\bibfnamefont {T.}~\bibnamefont {Hioki}},\
		and\ \bibinfo {author} {\bibfnamefont {Y.}~\bibnamefont {Masuda}},\
	}\bibfield  {title} {\bibinfo {title} {Hyperfine fields in an incommensurate
			antiferromagnetic {Cr-Mo} alloy system},\ }\href
	{https://doi.org/10.1143/JPSJ.39.672} {\bibfield  {journal} {\bibinfo
			{journal} {J. Phys. Soc. Jpn.}\ }\textbf {\bibinfo {volume} {39}},\ \bibinfo
		{pages} {672} (\bibinfo {year} {1975})}\BibitemShut {NoStop}%
	\bibitem [{\citenamefont {Sakurai}\ \emph {et~al.}(2002)\citenamefont
		{Sakurai}, \citenamefont {Tsuboi}, \citenamefont {Kato}, \citenamefont
		{Yoshimura}, \citenamefont {Kosuge}, \citenamefont {Mitsuda}, \citenamefont
		{Mitamura},\ and\ \citenamefont {Goto}}]{Sakurai024428}%
	\BibitemOpen
	\bibfield  {author} {\bibinfo {author} {\bibfnamefont {H.}~\bibnamefont
			{Sakurai}}, \bibinfo {author} {\bibfnamefont {N.}~\bibnamefont {Tsuboi}},
		\bibinfo {author} {\bibfnamefont {M.}~\bibnamefont {Kato}}, \bibinfo {author}
		{\bibfnamefont {K.}~\bibnamefont {Yoshimura}}, \bibinfo {author}
		{\bibfnamefont {K.}~\bibnamefont {Kosuge}}, \bibinfo {author} {\bibfnamefont
			{A.}~\bibnamefont {Mitsuda}}, \bibinfo {author} {\bibfnamefont
			{H.}~\bibnamefont {Mitamura}},\ and\ \bibinfo {author} {\bibfnamefont
			{T.}~\bibnamefont {Goto}},\ }\bibfield  {title} {\bibinfo {title}
		{Antiferromagnetic order in the two-dimensional spin system
			{Cu$_3$B$_2$O$_6$}},\ }\href {https://doi.org/10.1103/PhysRevB.66.024428}
	{\bibfield  {journal} {\bibinfo  {journal} {Phys. Rev. B}\ }\textbf {\bibinfo
			{volume} {66}},\ \bibinfo {pages} {024428} (\bibinfo {year}
		{2002})}\BibitemShut {NoStop}%
	\bibitem [{\citenamefont {Ranjith}\ \emph {et~al.}(2016)\citenamefont
		{Ranjith}, \citenamefont {Nath}, \citenamefont {Majumder}, \citenamefont
		{Kasinathan}, \citenamefont {Skoulatos}, \citenamefont {Keller},
		\citenamefont {Skourski}, \citenamefont {Baenitz},\ and\ \citenamefont
		{Tsirlin}}]{Ranjith014415}%
	\BibitemOpen
	\bibfield  {author} {\bibinfo {author} {\bibfnamefont {K.~M.}\ \bibnamefont
			{Ranjith}}, \bibinfo {author} {\bibfnamefont {R.}~\bibnamefont {Nath}},
		\bibinfo {author} {\bibfnamefont {M.}~\bibnamefont {Majumder}}, \bibinfo
		{author} {\bibfnamefont {D.}~\bibnamefont {Kasinathan}}, \bibinfo {author}
		{\bibfnamefont {M.}~\bibnamefont {Skoulatos}}, \bibinfo {author}
		{\bibfnamefont {L.}~\bibnamefont {Keller}}, \bibinfo {author} {\bibfnamefont
			{Y.}~\bibnamefont {Skourski}}, \bibinfo {author} {\bibfnamefont
			{M.}~\bibnamefont {Baenitz}},\ and\ \bibinfo {author} {\bibfnamefont {A.~A.}\
			\bibnamefont {Tsirlin}},\ }\bibfield  {title} {\bibinfo {title}
		{{Commensurate and incommensurate magnetic order in spin-1 chains stacked on
				the triangular lattice in
				${\mathrm{Li}}_{2}{\mathrm{NiW}}_{2}{\mathrm{O}}_{8}$}},\ }\href
	{https://doi.org/10.1103/PhysRevB.94.014415} {\bibfield  {journal} {\bibinfo
			{journal} {Phys. Rev. B}\ }\textbf {\bibinfo {volume} {94}},\ \bibinfo
		{pages} {014415} (\bibinfo {year} {2016})}\BibitemShut {NoStop}%
	\bibitem [{\citenamefont {Ranjith}\ \emph {et~al.}(2015)\citenamefont
		{Ranjith}, \citenamefont {Majumder}, \citenamefont {Baenitz}, \citenamefont
		{Tsirlin},\ and\ \citenamefont {Nath}}]{Ranjith024422}%
	\BibitemOpen
	\bibfield  {author} {\bibinfo {author} {\bibfnamefont {K.~M.}\ \bibnamefont
			{Ranjith}}, \bibinfo {author} {\bibfnamefont {M.}~\bibnamefont {Majumder}},
		\bibinfo {author} {\bibfnamefont {M.}~\bibnamefont {Baenitz}}, \bibinfo
		{author} {\bibfnamefont {A.~A.}\ \bibnamefont {Tsirlin}},\ and\ \bibinfo
		{author} {\bibfnamefont {R.}~\bibnamefont {Nath}},\ }\bibfield  {title}
	{\bibinfo {title} {{Frustrated three-dimensional antiferromagnet
				${\text{Li}}_{2}{\text{CuW}}_{2}{\text{O}}_{8}$: $^{7}\mathrm{Li}$ NMR and
				the effect of nonmagnetic dilution}},\ }\href
	{https://doi.org/10.1103/PhysRevB.92.024422} {\bibfield  {journal} {\bibinfo
			{journal} {Phys. Rev. B}\ }\textbf {\bibinfo {volume} {92}},\ \bibinfo
		{pages} {024422} (\bibinfo {year} {2015})}\BibitemShut {NoStop}%
	\bibitem [{\citenamefont {Yamada}\ and\ \citenamefont
		{Sakata}(1986)}]{Yamada1751}%
	\BibitemOpen
	\bibfield  {author} {\bibinfo {author} {\bibfnamefont {Y.}~\bibnamefont
			{Yamada}}\ and\ \bibinfo {author} {\bibfnamefont {A.}~\bibnamefont
			{Sakata}},\ }\bibfield  {title} {\bibinfo {title} {An analysis method of
			antiferromagnetic powder patterns in spin-echo {NMR} under external fields},\
	}\href {https://doi.org/10.1143/JPSJ.55.1751} {\bibfield  {journal} {\bibinfo
			{journal} {J. Phys. Soc. Jpn.}\ }\textbf {\bibinfo {volume} {55}},\ \bibinfo
		{pages} {1751} (\bibinfo {year} {1986})}\BibitemShut {NoStop}%
	\bibitem [{\citenamefont {Kikuchi}\ \emph {et~al.}(2000)\citenamefont
		{Kikuchi}, \citenamefont {Ishiguchi}, \citenamefont {Motoya}, \citenamefont
		{Itoh}, \citenamefont {Inari}, \citenamefont {Eguchi},\ and\ \citenamefont
		{Akimitsu}}]{Kikuchi2660}%
	\BibitemOpen
	\bibfield  {author} {\bibinfo {author} {\bibfnamefont {J.}~\bibnamefont
			{Kikuchi}}, \bibinfo {author} {\bibfnamefont {K.}~\bibnamefont {Ishiguchi}},
		\bibinfo {author} {\bibfnamefont {K.}~\bibnamefont {Motoya}}, \bibinfo
		{author} {\bibfnamefont {M.}~\bibnamefont {Itoh}}, \bibinfo {author}
		{\bibfnamefont {K.}~\bibnamefont {Inari}}, \bibinfo {author} {\bibfnamefont
			{N.}~\bibnamefont {Eguchi}},\ and\ \bibinfo {author} {\bibfnamefont
			{J.}~\bibnamefont {Akimitsu}},\ }\bibfield  {title} {\bibinfo {title} {Nmr
			and neutron scattering studies of quasi one-dimensional magnet
			{CuV$_2$O$_6$}},\ }\href {https://doi.org/10.1143/JPSJ.69.2660} {\bibfield
		{journal} {\bibinfo  {journal} {J. Phys. Soc. Jpn.}\ }\textbf {\bibinfo
			{volume} {69}},\ \bibinfo {pages} {2660} (\bibinfo {year}
		{2000})}\BibitemShut {NoStop}%
	\bibitem [{\citenamefont {Johnston}\ \emph {et~al.}(2005)\citenamefont
		{Johnston}, \citenamefont {Baek}, \citenamefont {Zong}, \citenamefont
		{Borsa}, \citenamefont {Schmalian},\ and\ \citenamefont
		{Kondo}}]{Johnston176408}%
	\BibitemOpen
	\bibfield  {author} {\bibinfo {author} {\bibfnamefont {D.~C.}\ \bibnamefont
			{Johnston}}, \bibinfo {author} {\bibfnamefont {S.-H.}\ \bibnamefont {Baek}},
		\bibinfo {author} {\bibfnamefont {X.}~\bibnamefont {Zong}}, \bibinfo {author}
		{\bibfnamefont {F.}~\bibnamefont {Borsa}}, \bibinfo {author} {\bibfnamefont
			{J.}~\bibnamefont {Schmalian}},\ and\ \bibinfo {author} {\bibfnamefont
			{S.}~\bibnamefont {Kondo}},\ }\bibfield  {title} {\bibinfo {title} {{Dynamics
				of Magnetic Defects in Heavy Fermion ${\mathrm{LiV}}_{2}{\mathrm{O}}_{4}$
				from Stretched Exponential $^{7}\mathrm{Li}$ NMR Relaxation}},\ }\href
	{https://doi.org/10.1103/PhysRevLett.95.176408} {\bibfield  {journal}
		{\bibinfo  {journal} {Phys. Rev. Lett.}\ }\textbf {\bibinfo {volume} {95}},\
		\bibinfo {pages} {176408} (\bibinfo {year} {2005})}\BibitemShut {NoStop}%
	\bibitem [{\citenamefont {Moriya}(1956)}]{Moriya23}%
	\BibitemOpen
	\bibfield  {author} {\bibinfo {author} {\bibfnamefont {T.}~\bibnamefont
			{Moriya}},\ }\bibfield  {title} {\bibinfo {title} {Nuclear magnetic
			relaxation in antiferromagnetics},\ }\href
	{https://doi.org/10.1143/PTP.16.23} {\bibfield  {journal} {\bibinfo
			{journal} {Prog. Theor. Phys.}\ }\textbf {\bibinfo {volume} {16}},\ \bibinfo
		{pages} {23} (\bibinfo {year} {1956})}\BibitemShut {NoStop}%
	\bibitem [{\citenamefont {Beeman}\ and\ \citenamefont
		{Pincus}(1968)}]{Beeman359}%
	\BibitemOpen
	\bibfield  {author} {\bibinfo {author} {\bibfnamefont {D.}~\bibnamefont
			{Beeman}}\ and\ \bibinfo {author} {\bibfnamefont {P.}~\bibnamefont
			{Pincus}},\ }\bibfield  {title} {\bibinfo {title} {Nuclear spin-lattice
			relaxation in magnetic insulators},\ }\href
	{https://doi.org/10.1103/PhysRev.166.359} {\bibfield  {journal} {\bibinfo
			{journal} {Phys. Rev.}\ }\textbf {\bibinfo {volume} {166}},\ \bibinfo {pages}
		{359} (\bibinfo {year} {1968})}\BibitemShut {NoStop}%
	\bibitem [{\citenamefont {Belesi}\ \emph {et~al.}(2006)\citenamefont {Belesi},
		\citenamefont {Borsa},\ and\ \citenamefont {Powell}}]{Belesi184408}%
	\BibitemOpen
	\bibfield  {author} {\bibinfo {author} {\bibfnamefont {M.}~\bibnamefont
			{Belesi}}, \bibinfo {author} {\bibfnamefont {F.}~\bibnamefont {Borsa}},\ and\
		\bibinfo {author} {\bibfnamefont {A.~K.}\ \bibnamefont {Powell}},\ }\bibfield
	{title} {\bibinfo {title} {{Evidence for spin-wave excitations in the
				long-range magnetically ordered state of a ${\mathrm{Fe}}_{19}$ molecular
				crystal from proton NMR}},\ }\href
	{https://doi.org/10.1103/PhysRevB.74.184408} {\bibfield  {journal} {\bibinfo
			{journal} {Phys. Rev. B}\ }\textbf {\bibinfo {volume} {74}},\ \bibinfo
		{pages} {184408} (\bibinfo {year} {2006})}\BibitemShut {NoStop}%
	\bibitem [{\citenamefont {Nath}\ \emph {et~al.}(2009)\citenamefont {Nath},
		\citenamefont {Furukawa}, \citenamefont {Borsa}, \citenamefont {Kaul},
		\citenamefont {Baenitz}, \citenamefont {Geibel},\ and\ \citenamefont
		{Johnston}}]{Nath214430}%
	\BibitemOpen
	\bibfield  {author} {\bibinfo {author} {\bibfnamefont {R.}~\bibnamefont
			{Nath}}, \bibinfo {author} {\bibfnamefont {Y.}~\bibnamefont {Furukawa}},
		\bibinfo {author} {\bibfnamefont {F.}~\bibnamefont {Borsa}}, \bibinfo
		{author} {\bibfnamefont {E.~E.}\ \bibnamefont {Kaul}}, \bibinfo {author}
		{\bibfnamefont {M.}~\bibnamefont {Baenitz}}, \bibinfo {author} {\bibfnamefont
			{C.}~\bibnamefont {Geibel}},\ and\ \bibinfo {author} {\bibfnamefont {D.~C.}\
			\bibnamefont {Johnston}},\ }\bibfield  {title} {\bibinfo {title}
		{{Single-crystal $^{31}\text{P}$ NMR studies of the frustrated square-lattice
				compound ${\text{Pb}}_{2}(\text{VO}){({\text{PO}}_{4})}_{2}$}},\ }\href
	{https://doi.org/10.1103/PhysRevB.80.214430} {\bibfield  {journal} {\bibinfo
			{journal} {Phys. Rev. B}\ }\textbf {\bibinfo {volume} {80}},\ \bibinfo
		{pages} {214430} (\bibinfo {year} {2009})}\BibitemShut {NoStop}%
	\bibitem [{\citenamefont {Matsubara}(1982)}]{Matsubara2424}%
	\BibitemOpen
	\bibfield  {author} {\bibinfo {author} {\bibfnamefont {F.}~\bibnamefont
			{Matsubara}},\ }\bibfield  {title} {\bibinfo {title} {{Magnetic Ordering in a
				Hexagonal Antiferromagnet}},\ }\href {https://doi.org/10.1143/JPSJ.51.2424}
	{\bibfield  {journal} {\bibinfo  {journal} {J. Phys. Soc. Jpn.}\ }\textbf
		{\bibinfo {volume} {51}},\ \bibinfo {pages} {2424} (\bibinfo {year}
		{1982})}\BibitemShut {NoStop}%
	\bibitem [{\citenamefont {Miyashita}\ and\ \citenamefont
		{Kawamura}(1985)}]{Miyashita3385}%
	\BibitemOpen
	\bibfield  {author} {\bibinfo {author} {\bibfnamefont {S.}~\bibnamefont
			{Miyashita}}\ and\ \bibinfo {author} {\bibfnamefont {H.}~\bibnamefont
			{Kawamura}},\ }\bibfield  {title} {\bibinfo {title} {{Phase Transitions of
				Anisotropic Heisenberg Antiferromagnets on the Triangular Lattice}},\ }\href
	{https://doi.org/10.1143/JPSJ.54.3385} {\bibfield  {journal} {\bibinfo
			{journal} {J. Phys. Soc. Jpn.}\ }\textbf {\bibinfo {volume} {54}},\ \bibinfo
		{pages} {3385} (\bibinfo {year} {1985})}\BibitemShut {NoStop}%
	\bibitem [{\citenamefont {Quirion}\ \emph {et~al.}(2015)\citenamefont
		{Quirion}, \citenamefont {Lapointe-Major}, \citenamefont {Poirier},
		\citenamefont {Quilliam}, \citenamefont {Dun},\ and\ \citenamefont
		{Zhou}}]{Quirion014414}%
	\BibitemOpen
	\bibfield  {author} {\bibinfo {author} {\bibfnamefont {G.}~\bibnamefont
			{Quirion}}, \bibinfo {author} {\bibfnamefont {M.}~\bibnamefont
			{Lapointe-Major}}, \bibinfo {author} {\bibfnamefont {M.}~\bibnamefont
			{Poirier}}, \bibinfo {author} {\bibfnamefont {J.~A.}\ \bibnamefont
			{Quilliam}}, \bibinfo {author} {\bibfnamefont {Z.~L.}\ \bibnamefont {Dun}},\
		and\ \bibinfo {author} {\bibfnamefont {H.~D.}\ \bibnamefont {Zhou}},\
	}\bibfield  {title} {\bibinfo {title} {{Magnetic phase diagram of
				${\mathrm{Ba}}_{3}{\mathrm{CoSb}}_{2}{\mathrm{O}}_{9}$ as determined by
				ultrasound velocity measurements}},\ }\href
	{https://doi.org/10.1103/PhysRevB.92.014414} {\bibfield  {journal} {\bibinfo
			{journal} {Phys. Rev. B}\ }\textbf {\bibinfo {volume} {92}},\ \bibinfo
		{pages} {014414} (\bibinfo {year} {2015})}\BibitemShut {NoStop}%
	\bibitem [{\citenamefont {Lee}\ \emph {et~al.}(2014)\citenamefont {Lee},
		\citenamefont {Hwang}, \citenamefont {Choi}, \citenamefont {Ma},
		\citenamefont {Dela~Cruz}, \citenamefont {Zhu}, \citenamefont {Ke},
		\citenamefont {Dun},\ and\ \citenamefont {Zhou}}]{Lee104420}%
	\BibitemOpen
	\bibfield  {author} {\bibinfo {author} {\bibfnamefont {M.}~\bibnamefont
			{Lee}}, \bibinfo {author} {\bibfnamefont {J.}~\bibnamefont {Hwang}}, \bibinfo
		{author} {\bibfnamefont {E.~S.}\ \bibnamefont {Choi}}, \bibinfo {author}
		{\bibfnamefont {J.}~\bibnamefont {Ma}}, \bibinfo {author} {\bibfnamefont
			{C.~R.}\ \bibnamefont {Dela~Cruz}}, \bibinfo {author} {\bibfnamefont
			{M.}~\bibnamefont {Zhu}}, \bibinfo {author} {\bibfnamefont {X.}~\bibnamefont
			{Ke}}, \bibinfo {author} {\bibfnamefont {Z.~L.}\ \bibnamefont {Dun}},\ and\
		\bibinfo {author} {\bibfnamefont {H.~D.}\ \bibnamefont {Zhou}},\ }\bibfield
	{title} {\bibinfo {title} {{Series of phase transitions and multiferroicity
				in the quasi-two-dimensional spin-$\frac{1}{2}$ triangular-lattice
				antiferromagnet ${\mathrm{Ba}}_{3}{\mathrm{CoNb}}_{2}{\mathrm{O}}_{9}$}},\
	}\href {https://doi.org/10.1103/PhysRevB.89.104420} {\bibfield  {journal}
		{\bibinfo  {journal} {Phys. Rev. B}\ }\textbf {\bibinfo {volume} {89}},\
		\bibinfo {pages} {104420} (\bibinfo {year} {2014})}\BibitemShut {NoStop}%
	\bibitem [{\citenamefont {Zhou}\ \emph {et~al.}(2012)\citenamefont {Zhou},
		\citenamefont {Xu}, \citenamefont {Hallas}, \citenamefont {Silverstein},
		\citenamefont {Wiebe}, \citenamefont {Umegaki}, \citenamefont {Yan},
		\citenamefont {Murphy}, \citenamefont {Park}, \citenamefont {Qiu},
		\citenamefont {Copley}, \citenamefont {Gardner},\ and\ \citenamefont
		{Takano}}]{Zhou267206}%
	\BibitemOpen
	\bibfield  {author} {\bibinfo {author} {\bibfnamefont {H.~D.}\ \bibnamefont
			{Zhou}}, \bibinfo {author} {\bibfnamefont {C.}~\bibnamefont {Xu}}, \bibinfo
		{author} {\bibfnamefont {A.~M.}\ \bibnamefont {Hallas}}, \bibinfo {author}
		{\bibfnamefont {H.~J.}\ \bibnamefont {Silverstein}}, \bibinfo {author}
		{\bibfnamefont {C.~R.}\ \bibnamefont {Wiebe}}, \bibinfo {author}
		{\bibfnamefont {I.}~\bibnamefont {Umegaki}}, \bibinfo {author} {\bibfnamefont
			{J.~Q.}\ \bibnamefont {Yan}}, \bibinfo {author} {\bibfnamefont {T.~P.}\
			\bibnamefont {Murphy}}, \bibinfo {author} {\bibfnamefont {J.-H.}\
			\bibnamefont {Park}}, \bibinfo {author} {\bibfnamefont {Y.}~\bibnamefont
			{Qiu}}, \bibinfo {author} {\bibfnamefont {J.~R.~D.}\ \bibnamefont {Copley}},
		\bibinfo {author} {\bibfnamefont {J.~S.}\ \bibnamefont {Gardner}},\ and\
		\bibinfo {author} {\bibfnamefont {Y.}~\bibnamefont {Takano}},\ }\bibfield
	{title} {\bibinfo {title} {{Successive Phase Transitions and Extended
				Spin-Excitation Continuum in the spin-$\frac{1}{2}$ Triangular-Lattice
				Antiferromagnet ${\mathrm{Ba}}_{3}{\mathrm{CoSb}}_{2}{\mathrm{O}}_{9}$}},\
	}\href {https://doi.org/10.1103/PhysRevLett.109.267206} {\bibfield  {journal}
		{\bibinfo  {journal} {Phys. Rev. Lett.}\ }\textbf {\bibinfo {volume} {109}},\
		\bibinfo {pages} {267206} (\bibinfo {year} {2012})}\BibitemShut {NoStop}%
	\bibitem [{\citenamefont {Ranjith}\ \emph {et~al.}(2017)\citenamefont
		{Ranjith}, \citenamefont {Brinda}, \citenamefont {Arjun}, \citenamefont
		{Hegde},\ and\ \citenamefont {Nath}}]{Ranjith115804}%
	\BibitemOpen
	\bibfield  {author} {\bibinfo {author} {\bibfnamefont {K.~M.}\ \bibnamefont
			{Ranjith}}, \bibinfo {author} {\bibfnamefont {K.}~\bibnamefont {Brinda}},
		\bibinfo {author} {\bibfnamefont {U.}~\bibnamefont {Arjun}}, \bibinfo
		{author} {\bibfnamefont {N.~G.}\ \bibnamefont {Hegde}},\ and\ \bibinfo
		{author} {\bibfnamefont {R.}~\bibnamefont {Nath}},\ }\bibfield  {title}
	{\bibinfo {title} {{Double phase transition in the triangular antiferromagnet
				${\mathrm{Ba}}_{3}{\mathrm{Co}}{\mathrm{Ta}}_{2}{\mathrm{O}}_{9}$}},\ }\href
	{https://doi.org/10.1088/1361-648x/aa57be} {\bibfield  {journal} {\bibinfo
			{journal} {J. Phys.: Condens. Matter}\ }\textbf {\bibinfo {volume} {29}},\
		\bibinfo {pages} {115804} (\bibinfo {year} {2017})}\BibitemShut {NoStop}%
	\bibitem [{\citenamefont {Yokota}\ \emph {et~al.}(2014)\citenamefont {Yokota},
		\citenamefont {Kurita},\ and\ \citenamefont {Tanaka}}]{Yokota014403}%
	\BibitemOpen
	\bibfield  {author} {\bibinfo {author} {\bibfnamefont {K.}~\bibnamefont
			{Yokota}}, \bibinfo {author} {\bibfnamefont {N.}~\bibnamefont {Kurita}},\
		and\ \bibinfo {author} {\bibfnamefont {H.}~\bibnamefont {Tanaka}},\
	}\bibfield  {title} {\bibinfo {title} {{Magnetic phase diagram of the
				spin-$\frac{1}{2}$ triangular-lattice Heisenberg antiferromagnet
				${\mathrm{Ba}}_{3}\mathrm{Co}{\mathrm{Nb}}_{2}{\mathrm{O}}_{9}$}},\ }\href
	{https://doi.org/10.1103/PhysRevB.90.014403} {\bibfield  {journal} {\bibinfo
			{journal} {Phys. Rev. B}\ }\textbf {\bibinfo {volume} {90}},\ \bibinfo
		{pages} {014403} (\bibinfo {year} {2014})}\BibitemShut {NoStop}%
	\bibitem [{\citenamefont {Rai}\ \emph {et~al.}(2006)\citenamefont {Rai},
		\citenamefont {Cao}, \citenamefont {Brown}, \citenamefont {Musfeldt},
		\citenamefont {Kasinathan}, \citenamefont {Singh}, \citenamefont {Lawes},
		\citenamefont {Rogado}, \citenamefont {Cava},\ and\ \citenamefont
		{Wei}}]{Rai235101}%
	\BibitemOpen
	\bibfield  {author} {\bibinfo {author} {\bibfnamefont {R.~C.}\ \bibnamefont
			{Rai}}, \bibinfo {author} {\bibfnamefont {J.}~\bibnamefont {Cao}}, \bibinfo
		{author} {\bibfnamefont {S.}~\bibnamefont {Brown}}, \bibinfo {author}
		{\bibfnamefont {J.~L.}\ \bibnamefont {Musfeldt}}, \bibinfo {author}
		{\bibfnamefont {D.}~\bibnamefont {Kasinathan}}, \bibinfo {author}
		{\bibfnamefont {D.~J.}\ \bibnamefont {Singh}}, \bibinfo {author}
		{\bibfnamefont {G.}~\bibnamefont {Lawes}}, \bibinfo {author} {\bibfnamefont
			{N.}~\bibnamefont {Rogado}}, \bibinfo {author} {\bibfnamefont {R.~J.}\
			\bibnamefont {Cava}},\ and\ \bibinfo {author} {\bibfnamefont
			{X.}~\bibnamefont {Wei}},\ }\bibfield  {title} {\bibinfo {title} {{Optical
				properties and magnetic-field-induced phase transitions in the ferroelectric
				state of ${\mathrm{Ni}}_{3}{\mathrm{V}}_{2}{\mathrm{O}}_{8}$: Experiments and
				first-principles calculations}},\ }\href
	{https://doi.org/10.1103/PhysRevB.74.235101} {\bibfield  {journal} {\bibinfo
			{journal} {Phys. Rev. B}\ }\textbf {\bibinfo {volume} {74}},\ \bibinfo
		{pages} {235101} (\bibinfo {year} {2006})}\BibitemShut {NoStop}%
	\bibitem [{\citenamefont {Wilson}\ \emph {et~al.}(2007)\citenamefont {Wilson},
		\citenamefont {Petrenko},\ and\ \citenamefont {Chapon}}]{Wilson094432}%
	\BibitemOpen
	\bibfield  {author} {\bibinfo {author} {\bibfnamefont {N.~R.}\ \bibnamefont
			{Wilson}}, \bibinfo {author} {\bibfnamefont {O.~A.}\ \bibnamefont
			{Petrenko}},\ and\ \bibinfo {author} {\bibfnamefont {L.~C.}\ \bibnamefont
			{Chapon}},\ }\bibfield  {title} {\bibinfo {title} {{Magnetic phases in the
				Kagom\'e staircase compound
				${\mathrm{Co}}_{3}{\mathrm{V}}_{2}{\mathrm{O}}_{8}$ studied using powder
				neutron diffraction}},\ }\href {https://doi.org/10.1103/PhysRevB.75.094432}
	{\bibfield  {journal} {\bibinfo  {journal} {Phys. Rev. B}\ }\textbf {\bibinfo
			{volume} {75}},\ \bibinfo {pages} {094432} (\bibinfo {year}
		{2007})}\BibitemShut {NoStop}%
	\bibitem [{\citenamefont {Dong}\ \emph {et~al.}(2022)\citenamefont {Dong},
		\citenamefont {Wang}, \citenamefont {He}, \citenamefont {Chang},
		\citenamefont {Shi}, \citenamefont {Song}, \citenamefont {Jin}, \citenamefont
		{Du}, \citenamefont {Wu}, \citenamefont {Han}, \citenamefont {Kindo},\ and\
		\citenamefont {Yang}}]{Dong024427}%
	\BibitemOpen
	\bibfield  {author} {\bibinfo {author} {\bibfnamefont {C.}~\bibnamefont
			{Dong}}, \bibinfo {author} {\bibfnamefont {J.~F.}\ \bibnamefont {Wang}},
		\bibinfo {author} {\bibfnamefont {Z.~Z.}\ \bibnamefont {He}}, \bibinfo
		{author} {\bibfnamefont {Y.~T.}\ \bibnamefont {Chang}}, \bibinfo {author}
		{\bibfnamefont {M.~Y.}\ \bibnamefont {Shi}}, \bibinfo {author} {\bibfnamefont
			{Y.~R.}\ \bibnamefont {Song}}, \bibinfo {author} {\bibfnamefont {S.~M.}\
			\bibnamefont {Jin}}, \bibinfo {author} {\bibfnamefont {Y.~Q.}\ \bibnamefont
			{Du}}, \bibinfo {author} {\bibfnamefont {Z.~Y.}\ \bibnamefont {Wu}}, \bibinfo
		{author} {\bibfnamefont {X.~T.}\ \bibnamefont {Han}}, \bibinfo {author}
		{\bibfnamefont {K.}~\bibnamefont {Kindo}},\ and\ \bibinfo {author}
		{\bibfnamefont {M.}~\bibnamefont {Yang}},\ }\bibfield  {title} {\bibinfo
		{title} {{Reentrant ferroelectric phase induced by a tilting high magnetic
				field in ${\mathrm{Ni}}_{3}{\mathrm{V}}_{2}{\mathrm{O}}_{8}$}},\ }\href
	{https://doi.org/10.1103/PhysRevB.105.024427} {\bibfield  {journal} {\bibinfo
			{journal} {Phys. Rev. B}\ }\textbf {\bibinfo {volume} {105}},\ \bibinfo
		{pages} {024427} (\bibinfo {year} {2022})}\BibitemShut {NoStop}%
\end{thebibliography}

\end{document}